\documentclass[11pt]{article}%[12pt]{iopart}
\begin{document} 
\title{Structure of Higher Spin Gauge Interactions}        
\author{Anders K. H. Bengtsson\footnote{e-mail: anders.bengtsson@hb.se. Work supported by the Research and Education Board at the University College of Bor{\aa}s.}}
\date{November 6, 2006}
\maketitle
\begin{center}School of Engineering\break University College of Bor{\aa}s\break All\' egatan 1, SE-50190 Bor\aa s, Sweden.\end{center}

\begin{abstract}
In a previous paper, higher spin gauge field theory was formulated in an abstract way, essentially only keeping enough machinery to discuss gauge invariance of an action. The approach could be thought of as providing an interface (or syntax) towards an implementation (or semantics) yet to be constructed. The structure then revealed turned out to be that of a strongly homotopy Lie algebra. 

In the present paper, the framework will be connected to more conventional field theoretic concepts. The Fock complex vertex operator implementation of the interactions in the BRST-BV formulation of the theory will be elaborated. The relation between the vertex order expansion and homological perturbation theory will be clarified. A formal non-obstruction argument is reviewed. The syntactically derived sh-Lie algebra structure is semantically mapped to the Fock complex implementation and it is shown that the equations governing the higher order vertices are reproduced. 

Global symmetries and subsidiary conditions are discussed and as a result the tracelessness constraints are discarded. Thus all equations needed to compute the vertices to any order are collected. The framework is general enough to encompass all possible interaction terms.

Finally, the abstract framework itself will be strengthened by showing that it can be naturally phrased in terms of  category theory.  
\end{abstract}
\pagebreak
\tableofcontents
\pagebreak

%===================================================================================================================
%INTRODUCTION
%===================================================================================================================
\section{Introduction}\label{sec:Introduction}
The present paper constitutes the second step in an attempt to rethink the BRST approach to higher spin gauge field theory \cite{AKHB1988}, the first one taken in \cite{AKHB2005a}, where we applied computer science inspired methods to the problem. 

While we would certainly expect conventional field theory tools from differential geometry, suitably generalized and elaborated, to be sufficiently powerful for a classical field theory of interacting higher spin gauge fields, it is not entirely clear what the proper generalizations might be, or whether any generalizations are indeed needed. As will be argued in the present paper, understanding the interaction vertices and finding explicit expressions for them might just be a question of devising a proper formalism which allows for efficient computation while keeping the inherent complexity of the problem at bay.

As a background to the ensuing discussion on formalism, consider the following train of thought. Employing the de Rahm complex, spin 1 gauge theory can be phrased in terms of 1-form vector fields ${\bf A}=A_{\mu}dx^{\mu}$. For spin 2, we need to introduce vielbeins ${\bf e}^i=e_{\mu}^idx^{\mu}$ with $i$ a tangent space index. This makes sense, and the vielbeins are needed anyway in order to introduce spinor fields. For higher spin, the formalism then suggests considering spin-$s$ fields ${\bf f}^{i_1\ldots i_{s-1}}=f_{\mu}^{i_1\ldots i_{s-1}}dx^{\mu}$ and concomitant fields $\bf w$ generalizing the connections $\omega_\mu^{ij}$ accompanying the vielbeins. The formalism then starts to look lopsided with one index being treated in a special way in order to make the theory generally covariant. This is not to say that the form language is not without its uses. As long as we are interested in local functionals of the fields, like the action itself, the form language is practical. However, this does not imply that the fields themselves must be taken as 1-forms. 

Still, the hitherto most succesful approach to interactions, developed by M. Vasiliev, is formulated within a differential geometric framework, basically using forms, or frame fields, as described above \cite{Vasiliev2004Reviewa,BekaertCnockaertIazeollaVasiliev2005}. The approach is generally covariant by construction. Strong arguments has been put forward supporting this approach based on the very special nature of the gravitational interaction and general relativity and the problems with coupling higher spin gauge fields to gravity. The Vasiliev approach provides non-linear field equations for various infinite tower multiplets of higher spin gauge fields depending on the underlying higher spin algebra. Furthermore, the approach supports quite a few connections to string theory related models where higher spin massless excitations occur (see \cite{FirstSolvayWorkshop} for reviews and extensive lists of references). It might then be asked, why pursue yet another approach?

As discussed in more detail in \cite{AKHB2005a} (where a list of original references to this complex of problems can be found), there are roughly two main approaches to introducing interactions, either gauging a global non-Abelian symmetry algebra, or deforming a free theory local Abelian gauge group. For spin 1 the two approaches give the same result, namely Yang-Mills theory. For spin 2, barring conceptual issues and technicalities, the results agree with Einstein gravity.

The Vasiliev approach to higher spin gauge field interactions could be considered to belong to the first ''gauging'' tradition as its constructions are built around the higher spin algebras, themselves extensions of the AdS algebra. By comparison, the BRST approach of the present paper and its precursors, is squarely in the second ''deformation'' tradition. If we could construct higher spin interactions in this approach we would be able to explicitly compare the result with that of Vasiliev. This is one motivation behind pursuing the BRST deformation approach. For free field theory this comparison has been done in \cite{BarnichGrigorievSemikhatovTipunin2004} by showing that both the unfolded formulation of Vasiliev and the BRST formulation can be derived from a so-called parent field theory.

The BRST formulation is based on symmetric tensor fields which from the point of view of an infinite tower of gauge fields is quite natural. Still, the inclusion of matter fields (which will not be discussed here) and half-integer higher spin gauge fields would require vielbeins. This might point towards a formulation purely in terms of spinors or twistors. This could be discussed at length, suffice it to say that the present paper is based on treating all integer spin gauge fields on a common ground with no special a priori assumptions about spin 2. This then entails that spacetime curvature, and gravitational interactions, will be an output of the theory rather than an input. The approach has to be judged based on its eventual success or failure.
 
In the previous publication \cite{AKHB2005a}, an abstract approach was proposed to higher spin gauge field theory which amounted to providing a syntactic interface (in the sense of computer science) to gauge fields and the action. The main purpose was to free the theory from, possibly delimiting, assumptions some of which might turn out to be circumstantial or arbitrary. Thus we discarded almost all of conventional field theory paraphernalia and just stated that all higher spin gauge fields can be packed away into abstract objects $\Phi$ of a certain type ${\cal H}$. For the free field theory we have at least one concrete implementation to fall back upon, namely the explicit $\langle\Phi|Q|\Phi\rangle$ action \cite{OuvryStern1987a,AKHB1987a,SiegelZwiebach1987,AKHB1988} in terms of a first quantized mechanics BRST formulation of the classical field theory. 

The abstract approach showed that interacting higher spin gauge field theory, when studied in a deformation theoretic context, must fall into the structure of strongly homotopy Lie algebra, for short a $L_\infty$-algebra \cite{LadaStasheff1993a}. Thus we were able to corroborate earlier deformation theoretic work. This result, it seems in retrospect, has been more or less implicit in similar approaches to higher spins, but the abstract approach shows that the result is inevitable in any formal power series approach. Indeed, the result is not really particular to higher spins per se, but rather to formal power series expansions of gauge invariant actions and concomitant gauge transformations.

The first systematic discussion of higher spin gauge interactions after \cite{FangFronsdal1979} (where the problem was explicitly formulated) and \cite{BBB1983a,BerendsBurgersvanDam1984} (where the first positive results were reported) can be found in the work of Berends, Burgers and van Dam \cite{BerendsBurgersvanDam1985,Burgers1985thesis} (BBvD). Although these authors did not explicitly treat an infinite tower of higher spin gauge fields, their formalism can be adapted to that case. Later on, Fulp, Lada and Stasheff proved that under the assumption that the gauge theory discussed by BBvD exists, more precisely that the gauge algebra considered by BBvD closes on an infinite tower of higher spin gauge fields (the {\it BBvD-hypothesis}) then the gauge algebra must be a $L_\infty$ algebra \cite{FulpLadaStasheff2002}. Barnich and Henneaux \cite{BarnichHenneaux1993a,Henneaux1997a} pointed out that this deformation theoretic approach to higher spin gauge interactions can be cast into the Batalin-Vilkovisky antifield formalism \cite{BatalinVilkovisky1985a}, but did not provide details. Another early paper discussing interactions in a BV framework is \cite{FougereKnechtStern1991}.

There are further indications \cite{BarnichFulpLadaStasheff1998,Stasheff1997a} that a formulation within the full variational bi-complex \cite{IMAndersson1992}, most likely augmented with BV field/antifield tools \cite{McCloud1994,Brandt1998}, might be powerful enough to deal with the complexities of interacting higher spin gauge fields. 

There are thus quite a few general schemes in the literature, set up in order to derive gauge invariant interacting field theories by deforming a free field theory \cite{BerendsBurgersvanDam1985,Anco1992,BarnichHenneaux1993a}. When it comes to implementing the schemes in practice, none of them goes beyond supergravity, perhaps with the caveat that recently, some new results on cubic interactions for higher spins have appeared (reviewed in \cite{BekaertBoulangerCnockaertLeclercq2006a}) that goes beyond the original papers \cite{BBB1983a,BerendsBurgersvanDam1984}. Again, the reason is of course the immense complexity which dwarfs even the most elaborate scheme, and which is indeed apparent from the recent work on cubic terms \cite{BekaertBoulangerCnockaert2006a} and new work \cite{BuchbinderFotopoulosPetkouTsulaia2006a} along the lines of the initial attempt in \cite{AKHB1988}. The question is precisely how to control the complexity. Speaking of complexity, it would indeed be interesting to probe further into the question of what complexity class the computation of general spin-$s$, $n$-point interaction vertices fall into.

Part of the motivation behind the present paper is to bring all these scattered strands together and try to focus them into a structured attempt to construct the interactions within a particular model. It should be stressed that we are interested in the {\it general structure} of higher spin gauge theory while considering as a running example only {\it the simplest model}.

That said, in the present work we will advance the project in a few directions. 

\begin{description}
\item{(i)}
We will connect the abstract approach more closely to conventional field theory, and in doing so we will switch focus from gauge invariance to field theory BRST invariance proper, i.e. using spin 1 as template, we will switch from considering gauge transformations $\delta A=d\xi$ to considering BRST transformations $\delta A=d{\cal C}$ with ${\cal C}$ the ghost field corresponding to the gauge parameter $\xi$. The implementation in terms of the mechanics Fock space complex will be recast into the BV field-antifield framework. This provides for technical advantages, yields some new insights, and perhaps it is also more fundamental.

\item{(ii)} 
Then we will elaborate the vertex implementation of the interacting theory within the $\langle\Phi|Q|\Phi\rangle$ formulation of the free theory, lifted to BV formalism according to item (i). Attention will be drawn to a few points, sometimes overlooked, such as global symmetries. Thus all equations governing the vertices to all orders will be collected for the first time. The purpose is mainly to prepare the ground for a full scale computerized attack on the first few orders of interaction vertices. To keep it simple we will have just one primary field at each level of spin, but make no assumption about spacetime dimension. Unless the approach fails in the simplest case, there is really nothing to be gained from increasing the complexity by considering more complicated multiplets of fields. 

\item{(iii)} 
On the formal side, the sh-Lie algebra structure, syntactically derived within the abstract approach, will be semantically mapped to the Fock complex implementation. The product identities of the $L_\infty$ algebra then maps precisely into the equations governing the vertices to all orders. The abstract approach itself will be briefly discussed in terms of category theoretic concepts such as operads. This is important not just for understanding the structure of the classical field theory, but also for understanding the problems of quantization that will eventually ensue, given that the classical field theory can be constructed. One reason why category theory might be relevant in the present context is the simple observation that category theory allow us to discuss structure without having to deal with low level machinery.

\end{description}

Thus, in section \ref{sec:Fields} we will reexamine the basic assumptions of higher spin theory, establish notation and discuss some items of intuition guiding the subsequent development. An explicit higher spin fiber, to be used in all subsequent low level modeling, is defined.

In section \ref{sec:BRSTBV}, the connection between the first quantized mechanics BRST theory and the field theoretic BV antifield - antibracket formalism will be reviewed while adapting the formalism to the problem at hand. 
In section \ref{sec:DeformationTheory}, we discuss at length the relevant deformation theory, in particular the question of obstructions to higher spin gauge interactions. The overall structure of the cubic and quartic interactions are investigated and then the equations governing interactions to all orders are derived. The sh-Lie structure is revealed. Section \ref{sec:BRSTBV} and \ref{sec:DeformationTheory} then together provide a thorough discussion of higher spin gauge field theory in the BRST-BV formalism.

In section \ref{sec:VertexImplementation}, the Fock complex vertex implementation (based on the previously introduced higher spin fiber) is developed further to the point that the next step is to write explicit code to compute the first few orders in interaction vertices. We also clarify the role played by global symmetries and the tracelessness constraints. We argue, based on counting the number of physical states, that the tracelessness constraints can in fact be discarded in the BRST approach to the free theory due to the presence of the extra propagating components. As a consequence, apart from the primary gauge field at each level of spin, there will be propagating components of lower spin issuing from higher levels.

In section \ref{sec:Categories}, elements of a category theoretic formulation of the theory is sketched. The paper concludes with a conjecture as to the nature of the interactions in quantum theory.

Conventions, terminology and formulas that are relevant are collected in an appendix so as not to interrupt the line of argument in the main text. This should also enable connections to be made to the large and scattered literature pertaining to the subject at hand.

Lastly, it should be noted that higher spin gauge field theory pushes classical field theory to its limits. In sharp contrast to this, parts of the present paper might seem pedestrian, even elementary in places. The purpose is however to bring basic field theoretic assumptions to the fore, while at the same time gaining intuition and applying a principle of ''formalistic parsimony'', not letting us being drawn into details that cannot possibly affect the general solution to the interaction problem. Parsimony (see for example \cite{Gauch}) is a mainly empirical scientific method, which when transfered to theoretical science, could perhaps be thought of as ''instead of performing ten times as many new calculations, you look at what you already have and try to find the answer there''. We will thus lean on all sources and techniques we know off that might be relevant with the ostensible purpose off constructing a comprehensible and well structured approach to higher spin gauge field theory. The impatient reader can however skip at once to section \ref{sec:DeformationTheory} after studying table \ref{tab:Fields}.

Perhaps it should also be noted that we only discuss free-standing higher spin field theory. We will have nothing to say on other contexts where higher spin gauge fields occur.

As regards the terminology, we will speak of the "spin" of a gauge field as if we were really situated in four-dimensional spacetime, where by the way, we perhaps ought to use the term "helicity" instead, labeling the little group representations. Hopefully, this leisurely use of terminology will cause no confusion. By a spin $s$ field we simply mean a fully symmetric tensor field $\varphi_{\mu_1\mu_2\ldots\mu_s}$. Our concrete implementational model is easiest to set up in Minkowski background, but the general framework does not presuppose any particular spacetime background.

\pagebreak
%===================================================================================================================
%THE FIELDS
%===================================================================================================================
\section{The fields}\label{sec:Fields}
The problem of introducing interactions for higher spin gauge fields is easy to state but hard to solve. Involving as it does an infinite collection of fields and higher derivatives on fields, there is simply a problem of unlimited complexity. Then there is a lack of underlying principles reflecting a lack of understanding of their, purely conjectural of course, physical status. Although a descent into index purgatory might not be possible to put off indefinitely, a well tailored high-level formalism will certainly help to control the complexity. Ideally such a formalism should be based on underlying physical principles. These we do not have at the present, so we have to proceed tentatively. 

The conventional starting point is free fields on a fixed spacetime background. The mathematical framework becomes by default differential geometry. The purpose of the present section is to reexamine this framework while establishing our notation and to connect it to the abstract approach of \cite{AKHB2005a}. 

%**********
%THE FIELDS
%***********************************************
\subsection{Some basic facts}\label{subsec:SBF}

Let us start by reexamining the way we treat fields in mathematical physics, in particular higher spin gauge fields, with a backdrop to the following well known facts about higher spin gauge self-interactions.
\begin{description}
\item (1) Going beyond spin 2 requires an infinite tower of higher spin gauge fields to be introduced \cite{Fronsdal1979conf,AKHB1985,Burgers1985thesis}.
\item (2) The interaction vertices involve arbitrarily high order of spacetime derivatives \cite{BBB1983a,BerendsBurgersvanDam1984}. Power counting yields the minimal number of derivatives for an $n$-point, pure spin $s$ vertex to be $(n-2)s-2n+6$.
\item (3) Odd spin fields carry an internal index that is anti-symmetrized in the cubic interaction terms \cite{BBB1983a,BerendsBurgersvanDam1984}.
\item (4) The mass dimension of the coupling constant is $1-s$, where $s$ is the spin of the gauge field. This follows from power counting and item (2).
\end{description}

Thus it is clear that we need to consider from the outset an infinite collection of fields and derivatives of fields. In order to discuss such a situation without at once descending into details we need efficient notation expressing some common field theoretic rock bottom understanding.

%**********
%THE FIELDS
%***********************************************************
\subsection{What precisely is a field?}\label{subsec:WPIAF}
What is the most basic intuition behind the concept of a field in dynamics? First we must focus what we mean by dynamics. By a dynamical system we mean set of degrees of freedom $\{q_i\}$ whose time evolution is governed by a set of equations of the motion. There might also be constraints among the degrees of freedom. Denoting time by $t$, discrete or continuous, we would write $q_i(t)$. This notation is very flexible, as is certainly well known. We can think of $q_i(t)$ as symbolizing a particular degree of freedom labeled by $i$ at a certain time $t$, or we can think of it as the full collection of degrees of freedom at time $t$, or we can think of it as the complete history of either a particular degree of freedom or of the full collection. When we take the history point of view, we implicitly suppose that the $q_i$'s are solutions to equations of motion $W_i(q)=0$. These equations of motion in their turn might result from varying an action $A[q_i]$. In that case we think of the $q_i$'s as variables ranging over some huge space, and the histories are in that case generalized surfaces in that space satisfying the equations of motion. Both the degrees of freedom and the time parameter are in general subject to various symmetry transformations. 

This versatile nature of the mathematical notation can be confusing, and in this respect the language of mathematics is not that different from natural language. This versatility and conceptual vagueness of the notation mostly works to our advantage though.

Ordinary field theory works with fields, $\varphi^i(x)$, say. This notation can have various interpretations in different contexts. If $x$ is a "time" evolution parameter, we normally don't speak of field theory at all, but rather about "mechanics" and the $\varphi^i$ become generalized coordinates. This is the case already discussed. If $x$ are the coordinates of a spacetime, or of a space manifold ${\bf M}$, then we might have a field theory proper. As regards the index $i$ on the field, in one huge class of theories, it is thought of as coordinating a vector space. Why a vector space? Because without the ability to add fields and multiply them with numbers, there is not much we can do. Further mathematical elaborations lead to the concept of fields as sections of tensor bundles over a base manifold. This is wide enough to incorporate the index $i$ as collectively denoting spacetime tangent and cotangent indices $\mu,\nu,\ldots$ as well as internal structure group indices $a,b,\ldots$.

Since all these kinds of indices come with a vector space structure, they can be tensored, therefore it indeed makes sense to think of the fields as valued in a generalized vector space ${\bf H}$ parameterized collectively by $i$. Alternatively, the index $i$ could be thought of as a curved space index as in general relativity. Then we loose the vector space structure unless we reinstate it through (co)tangent spaces.

Again abstracting from the details, we see that field theory concerns itself with functions from one set ${\bf M}$ (with additional manifold structure), to another set ${\bf H}$ (with additional algebraic structure)

\begin{equation}\label{eq:FieldType}
f: {\bf M} \rightarrow {\bf H}
\end{equation}

or, more appropriately, the fiber bundle version of this. For that, denote the bundle with $({\bf E}, \pi, {\bf M})$. Fields considered as sections of $\pi$ are maps

\begin{equation}\label{eq:BundleType}
\varphi: {\bf M} \rightarrow {\bf E}
\end{equation}

satisfying $\pi\circ\varphi={\rm id}_M$. Locally, at least, we would have ${\bf E}={\bf M}\otimes{\bf H}$. For the purpose of formulating the theory, a globally trivial ${\bf E}$ is generally assumed. Again, attaching additional structure to the fibers makes all the difference. In the case of higher spin gauge fields we don't know what the fiber is and what that structure ought to be, that is indeed part of the problem. Most likely there is a collection of possible higher spin fibers. As already stated, we will only consider the simplest possible non-trivial higher spin fiber.

One thing that we do know of possible higher spin fibers is that they must be infinite dimensional since they must accommodate an infinite tower of tensor fields. Furthermore, since we know that the interactions will involve arbitrarily high derivatives, it is natural to think of the fiber as an infinite jet space $j^\infty{\bf E}$. Simply stated, this means that the field $\varphi$ itself and all its derivatives $\partial_{\mu_1}\partial_{\mu_2}\cdots\partial_{\mu_k}\varphi$ up to some finite but arbitrary order $k$ are considered as independent degrees of freedom \cite{Saunders}. We will however not explicitly employ such a framework in the present work. 

A further consideration is the nature of the background spacetime {\bf M}. For simplicity, we may take it as a fixed spacetime, and in that case, the whole bundle $\pi : {\bf E} \rightarrow {\bf M}$, is fixed. However, if the theory exists, then most likely there will be a back reaction on the full space ${\bf E}$ just as when gravity is constructed as a deformation of free spin 2 gauge theory. This is of course beyond what can be discussed presently, so for the time being the spacetime and fiber background will be fixed and trivial. This essentially turns the problem into an algebraic one.

It is often convenient to use DeWitt's condensed notation \cite{BSDeWitt1964,BSDeWitt1967b} in which all indices and spacetime dependence is package into one index $i$ and the fields are denoted by $\varphi^i$. If it is interesting to highlight the dependence on the base manifold, we can write $\varphi^i(x)$. In the first case, the abstract index $i$ belongs to a discrete set of indices as well as the base manifold continuous coordinates $x$. 

To conclude, with $x$ in {\bf M} and $i$ in {\bf H} we simply write a generic field as $\varphi^i(x)$. Thus the bundle approach to field theory seem to build in an asymmetry in the treatment of the base manifold and the fiber. A further abstraction comes about when we realize that there need be no distinction between the roles played by the base manifold and the fiber. This can be acknowledged by writing $\varphi(x,i)$, a notation that suggests a bilocal field theory based on an underlying mechanics model involving apart from $x\in{\bf M}$ also internal coordinates $i\in{\bf H}$. A further comment on this will appear below. It is clear that in this abstract notation, not very much is assumed about the mathematical structures to which $x$ and $i$ belong. 

%**********
%THE FIELDS
%*********************************************
\subsection{Abstract fields}\label{subsec:AF}
Let us now connect to the abstract interface of \cite{AKHB2005a}. The full collection of higher spin gauge fields is packaged into one object denoted by $\Phi$. The detailed packaging belongs to the implementation and will not concern us now. Most likely there are many different implementations. There are no a priori assumptions as to implementation in terms of fiber bundles over any manifold. When we need to distinguish different fields this will be done by writing $\Phi(\sigma_i)$ sometimes abbreviated to $\Phi_i$ for convenience. We see that we here actually build in one piece of classical field theory. Fields depend on variables, in general spacetime coordinates and possibly extra variables. All these are now collected into one indexed symbol $\sigma_i$, which need not be related to spacetime at all. Of course, this notation is open to all the ambiguities noted above. For now, the labeling can be thought of as just providing a handle to different instances of the field.

Indeed there seems to be two complementary ways to view an object such as $\Phi_i$. Either $i$ symbolizes a collection of variables (discrete and/or continuous) as in the DeWitt formalism, or it just symbolizes a handle to different instances of the class (or type) to which the particular object $\Phi_i$ belongs. This complementarity could be seen as analogous to the $x$-space vs. $p$-space complementarity in ordinary field theory.

We will work with two parallel ways of thinking about the abstract fields $\Phi$, one computer science oriented, the other mathematical. Thus we will consider an abstract data type $\cal H$ to which the fields $\Phi$ belong. This will be written $\Phi :: {\cal H}$ saying the $\Phi$ is of type $\cal H$. Particular, or generic, values of this type will be denoted by $\Phi(\sigma_i)$ or $\Phi_i$. Now a type always comes with a set of operations that can be performed on it. Eventually $\cal H$ will be implemented in a vector space or a Hilbert space. The operations supported are conveniently collected by stating that the type $\cal H$ supports the operations of a, possibly graded, vector space over $\bf C$, or some similar statement borrowing from mathematical language. Conforming more closely to computer science mode of expression, we would make a list of allowed operations on the type $\cal H$ and the corresponding types of the operations.

Thus $\cal H$ can be thought of as the abstraction of $\bf H$ which is the parallel mathematical meaning of the fields $\Phi$.

%**********
%THE FIELDS
%***********************************************************
\subsection{A simple higher spin fiber}\label{subsec:ASHSF}
Consider the fiber in the higher spin case. The fiber becomes infinite dimensional, parameterized in the simplest case by the helicity $\lambda=0,1,2,3,\ldots$ of the fields. In the first-quantized BRST approach we implement it as a Fock complex spanned by harmonic oscillators $\alpha^{\dagger}_\mu$ and accompanying ghost creators $c^+$ and $b_-$ acting freely on a doubly degenerate ground state $\{|+\rangle,|-\rangle\}$. We work in Minkowski spacetime for simplicity. For easy reference we just list the data needed to construct the fiber.

\begin{description}
\item{\it First quantized brackets}
\begin{equation}\label{eq:FirstQuantizedBrackets}
[\alpha_{\mu},\alpha_{\nu}^{\dagger}]=\eta_{\mu\nu},\quad\{c^+,b_+\}=\{c^-,b_-\}=\{c^0,b_0\}=1\label{eq:CommRels}.
\end{equation}

\item{\it Ghost hermitean conjugation}
\begin{equation}\label{eq:GhostConjugation}
(c^-)^\dagger=c^+,\quad(b_-)^\dagger=b_+,\quad(c^0)^\dagger=c^0,\quad(b_0)^\dagger=b_0.
\end{equation}

\item{\it Ground state properties}\label{eq:VacuumProperties}
\begin{eqnarray}
\alpha_\mu|+\rangle=\alpha_\mu|-\rangle=0,\nonumber \\
c^-|+\rangle=c^-|-\rangle=0, \nonumber\\
b_+|+\rangle=b_+|-\rangle=0, \nonumber\\
b_0|+\rangle=0,\quad b_0|-\rangle=|+\rangle,\\
c^0|-\rangle=0,\quad c^0|+\rangle=|-\rangle, \nonumber\\
\langle+|-\rangle=\langle-|+\rangle=1, \nonumber\\
\langle+|+\rangle=\langle-|-\rangle=0.\nonumber
\end{eqnarray}

\pagebreak
\item{\it Mass dimensionalities $d(\cdot)$}
\vskip 0.2cm
\begin{tabular}{lc}
Objects & $d(\cdot)$  \\ \hline
$p_\mu,c^+,c^-$ & 1 \\
$x_\mu,b^+,b_-$ & -1 \\
$\alpha_\mu,\alpha^\dagger_\mu,c^0,b_0$ & 0 \\ 
$|+\rangle,|-\rangle$ & 0\\
\end{tabular}
\begin{equation}\end{equation}

\item{\it Ghost number, $\rm gh_m(\cdot)$, and Grassman parity $\rho(\cdot)$ assignments}
\vskip 0.2cm
\begin{tabular}{lcc}
Objects & $\rm gh_m(\cdot)$ & $\rho(\cdot)$ \\ \hline
$\alpha_\mu,\alpha^\dagger_\mu$ & 0 & 0\\ 
$c^0,c^+,c^-$ &1 & 1\\ 
$b_0,b_+,b_-$ & -1 & 1\\ 
$|+\rangle$ & -1/2 & 0\\
$|-\rangle$ & 1/2 & 1\\ 
\end{tabular}
\begin{equation}\end{equation}

\item{\it Ghost complex structure}
\vskip 0.2cm
\begin{tabular}{ccccc}
$\rm gh_m(\cdot)$ & 3/2 & 1/2 & -1/2 & -3/2 \\ \hline
 & & $|-\rangle$ & $|+\rangle$ & \\ 
 & $c^+|-\rangle$ & $c^+|+\rangle$ & $c^+b_-|+\rangle$ & $b_-|+\rangle$ \\ 
  & & $c^+b_-|-\rangle$ & $b_-|-\rangle$ &  \\ 
\end{tabular}
\begin{equation}\end{equation} 

\item{\it Ghost complex expansion of physical fields}
\begin{equation}\label{eq:PhysicalFieldExpansion}
|\Phi\rangle=(A+Fc^+b_-+Hb_-c^0)|+\rangle.
\end{equation}

\item{\it Oscillator expansion of tensor fields}
\begin{equation}\label{eq:GenericOscillatorExpansion}
\phi=\phi_0+\phi^\mu\alpha^\dagger_\mu+\phi^{\mu\nu}\alpha^\dagger_\mu\alpha^\dagger_\nu+\ldots\;.\label{eq:FieldOscillatorExpansion}
\end{equation}

\item{\it Mechanics BRST operator}
\begin{equation}\label{eq:BRSToperator}
Q=-{1\over 2}c^0p^2+c^ + \alpha\cdot p + c^-\alpha^{\dagger}\cdot p + 2c^+c^-b_0.
\end{equation}

\end{description}

The actual construction of the fiber based on this data can be found in table \ref{tab:Fields} in section \ref{subsec:AISBVF}.

The physical basis of the fiber is an underlying two-particle mechanics model \cite{AKHB1987b}, slightly generalizing the various particle and superparticles considered in the literature. It is intuitively clear that if a field theory involving an infinite tower of fields is to be based on an underlying mechanics model, then this mechanics must be some kind of two-particle system with external spacetime translational movement and internal oscillations. Models of this kind were investigated in the 1960's in the connection to strong interaction physics, seemingly as a parallel track to S-matrix theory and dual models (see several papers in \cite{EighthNobelSymposium}). Most likely there are interesting connections to bilocal field theory to investigate (which was also done at the time), although of course, our present focus is entirely different.

\pagebreak
%===================================================================================================================
%MECHANICS BRST VS. BV FORMALISM
%===================================================================================================================
\section{Mechanics BRST vs. field theory BV formalism}\label{sec:BRSTBV}
If there is a mechanics model underlying a certain field theory, then one might suspect that there is some relation between the mechanics BRST operator and the field theory BRST transformations. Indeed there is, as was pointed out by Siegel in the context of string field theory \cite{Siegel1989}. Using the language of the Batalin-Vilkovisky field/antifield formalism \cite{BatalinVilkovisky1985a}, the correspondence is very close. We will now work out this connection explicitly for higher spin gauge fields.  As for the BV theory, we will use almost standard notation, thus the physical fields $\varphi^i$ and the ghosts $C^\alpha$ (and possibly ghosts for ghosts et cetera) are collectively denoted by $\Psi^A\supset\{\varphi^i,C^\alpha,\ldots\}$ and the corresponding antifields by $\Psi^\#_A\supset\{\varphi^\#_i,C^\#_\alpha,\ldots\}$.

For any functions $X$ and $Y$ of the fields/antifields the antibracket is defined as

\begin{equation}\label{eq:AntiBracket}
(X,Y)={\partial_r X\over\partial\Psi^A}{\partial_l Y\over\partial\Psi^\#_A}-{\partial_r X\over\partial\Psi^\#_A}{\partial_r Y\over\partial\Psi^A}.
\end{equation}

This is the standard definition, and under this bracket the fields/antifields are conjugate according to

\begin{equation}\label{eq:FieldAntiFieldBracket}
(\Psi^A,\Psi^\#_B)=\delta^A_B.
\end{equation}

The rest of the relevant formalism is collected in the appendix for easy reference, and we follow standard review literature closely \cite{HenneauxTeitelboim,Henneaux1990,BarnichBrandtHenneaux2000,GomisParisSamuel}.

In terms of the field theory BRST symmetry operator ${\bf s}$, the BRST transformations are written

\begin{equation}\label{eq:ActingS}
\delta_sX={\bf s}X,
\end{equation}

where we have suppressed an infinitesimal parameter that plays no role in the subsequent algebraic treatment.

As usual, $\bf s$ is an odd nilpotent derivation

\begin{equation}\label{eq:GrassmanS}
\rho({\bf s}X)=\rho_X+1 \quad({\rm mod}\;2),
\end{equation}

\begin{equation}\label{eq:DerivationS}
{\bf s}(XY)=X({\bf s}Y)+(-1)^{\rho_Y}({\bf s}X)Y,
\end{equation}
{
\begin{equation}\label{eq:NilpotencyS}
{\bf s}{\bf s}=0.
\end{equation}

The field theory ghost number grading is denoted by ${\rm gh_f}$ to disambiguate it from the mechanics ghost number ${\rm gh_m}$, and we have

\begin{equation}\label{eq:GhostNumberS}
{\rm gh_f}({\bf s}X)={\rm gh_f}(X)+1.
\end{equation}

Finally, we note that we want to implement ${\bf s}$ as a ''canonical'' transformation $S$ using the BV field/antifield bracket 

\begin{equation}\label{eq:CanonicalS}
{\bf s}X=(X,S).
\end{equation}

The right action of {\bf s} in (\ref{eq:DerivationS}) is chosen as to conform to this canonical action. For consistency, $S$ is even and has ghost number zero.

To summarize, ${\bf s}$ is a differential in a graded vector space, and cohomological groups can be defined as

\begin{equation}\label{eq:CohomologyS}
H^k({\bf s})=\Big({{\rm Ker}\;{\bf s}\over {\rm Im}\;{\bf s}}\Big)^k,
\end{equation}

where $H^0({\bf s})$ contains the gauge invariant functions at ghost number zero.

Next, we will implement this general structure explicitly within the Fock complex formulation of higher spin gauge theory of section \ref{subsec:ASHSF}.

%*******************************
%MECHANICS BRST VS. BV FORMALISM
%***************************************************************
\subsection{All integer spin BV formalism}\label{subsec:AISBVF}
The physical higher spin gauge fields (potentials, auxiliaries and traces of potentials) lie in the $\rm gh_m=-1/2$ sector, corresponding to equation (\ref{eq:PhysicalFieldExpansion}). The other three sectors can be naturally interpreted as antifields, ghosts and antighosts according to the following scheme. Here, we utilize the other-\linebreak wise abstract indices $A,B,C,\ldots$ to set up a convenient correspondence to fields/ghosts $\Psi^A$ and antifields/antighosts $\Psi^\#_A$. 

\vskip0.2cm
\begin{table}[h]
\begin{tabular}{llllllll}\label{tab:Fields}
Type 	    &Symbol	   &${\rm gh_m}$ &${\rm gh_f}$ &${\rm gh_a}$ &$\rho$ &d  &Field content \\ \hline
ghost     &$\Psi^1$  &-3/2         &\,\,1 	 &0 &1          &-1 &$|{\cal C}\rangle={\cal C}b_-|+\rangle$ \\
field     &$\Psi^0$  &-1/2 	     &\,\,0 	 &0 &0          &1  &$|\Phi\rangle=(A+Fc^+b_-+Hb_-c^0)|+\rangle$ \\
antifield &$\Psi^\#_0$ &\,\,1/2    &-1 		 &1 &1       &3 &$|\Phi^\#\rangle=(A^\#c^0+F^\#c^+b_-c^0+H^\#c^+)|+\rangle$ \\
antighost &$\Psi^\#_1$ &\,\,3/2    &-2           &2 &0 	        &5  &$|{\cal C}^\#\rangle={\cal C}^\#c^+c^0|+\rangle$ \\
\end{tabular}
\caption{Collection of data for the fields}
\label{tab:Fields}
\end{table}
\vskip0.2cm 

Field theory ghost number ${\rm gh_f}$ and mechanics ghost number ${\rm gh_m}$ are related through the formula ${\rm gh_f}=-({\rm gh_m}+1/2)$, i.e. compensate for the ghost number of the mechanics vacuum $|+\rangle$ and switch sign. Adding the antighost number ${\rm gh_a}$ and the ghost number ${\rm gh_f}$, yields the pure ghost number ${\rm gh_p}$ which is non-zero only for $\cal C$. $\rho$ is the Grassman parity. $d$ is the mass dimension of the $x$-space ket-fields. All Fock complex components are expanded into symmetric tensors according to the generic expansion in equation (\ref{eq:GenericOscillatorExpansion}). Note that $A$  contains the pure higher spin symmetric tensor fields. $H$ contains certain auxiliary fields which in the free theory are combinations of derivatives and divergencies of the higher spin symmetric tensors. $F$ contains a priori independent degrees of freedom unless certain constraints are applied in which case they are identified with traces of the higher spin symmetric tensors. This will be discussed in detail in sections \ref{subsec:GSASC} and \ref{subsec:DTC}.

There is one point regarding the Grassman properties of the fields to note here. With the vacuum $|+\rangle$ chosen to be even, all the Fock complex fields $|{\cal C}\rangle,|\Phi\rangle,|\Phi^\#\rangle$ and $|{\cal C}^\#\rangle$ become even. It is quite interesting that the mechanics BRST Grassman assignments precisely conspire with the BV field/antifield Grassman assignments in this way, and it is very convenient in calculations. Hence no Grassman signs occur when manipulating the bra and ket fields in table \ref{tab:Fields}. Note therefore that the $\rho$ in table \ref{tab:Fields} refers to the $\Psi$'s, not the $|\Psi\rangle$'s.

Note that the only data that is special in this table is the actual expansion of the fields in the particular Fock complex higher spin fiber, everything else perfectly general. For instance, the spectrum of ghost numbers $-2,-1,0,1$ is common to any irreducible Lagrangian gauge system \cite{LyakhovichSharapov2004}. Thus, as long as we do not expand out the components, our discussion is quite general.

In this representation, we expect to be able to implement the field theory BRST transformations in terms of the mechanics BRST operator $Q$. Indeed we find

\begin{eqnarray}\label{eq:BRSTTransformations1}
\delta|{\cal C}\rangle &=& O, \nonumber \\
\delta|\Phi\rangle &=& Q|{\cal C}\rangle, \nonumber \\
\delta|\Phi^\#\rangle &=& Q|\Phi\rangle, \nonumber \\
\delta|{\cal C}^\#\rangle &=& Q|\Phi^\#\rangle. \nonumber \\
\end{eqnarray}

This can also be written as a sequence

\begin{equation}\label{eq:ExactSequenceQ}
0\longrightarrow |{\cal C}\rangle\;{\stackrel{Q}{\longrightarrow}}\;|\Phi\rangle\;{\stackrel{Q}{\longrightarrow}}\;|\Phi^\#\rangle\;{\stackrel{Q}{\longrightarrow}}\;|{\cal C}^\#\rangle\;{\stackrel{Q}{\longrightarrow}}\;0,
\end{equation}

where we let the fields also denote the corresponding spaces of fields. In this sequence we find the gauge invariant physical fields in $H^0(Q)$ as

\begin{equation}\label{eq:QCohomology0}
Q|\Phi\rangle=0\quad{\rm mod}\quad |\Phi'\rangle=Q|{\cal C}\rangle,
\end{equation}

that is, precisely the space 

\begin{equation}\label{eq:KerQImQ0}
\Big({{\rm Ker}\;Q\over{\rm Im}\;Q}\Big)^0.
\end{equation}

By formally summing over ghost number, we can assemble these spaces of fields into one graded space (really an algebra)  $|\Psi\rangle$

\begin{equation}\label{eq:FormalPsi}
|\Psi\rangle=|{\cal C}\rangle\oplus|\Phi\rangle\oplus|\Phi^\#\rangle\oplus|{\cal C}^\#\rangle,
\end{equation}

where again we let $|\cdot\rangle$ do double work denoting both fields and spaces of fields.

The free field theory action now becomes a sum of two terms

\begin{equation}\label{eq:FreeS1}
S_0=S_0^{(0)}+S_0^{(1)}={1\over 2}\langle\Phi|Q|\Phi\rangle+\langle\Phi^\#|Q|{\cal C}\rangle,
\end{equation}

where the decoration ''$^{(i)}$'' denotes antighost number ${\rm gh_a}$. 

These are the only two bilinear terms of ghost number zero that can be formed out of fields/ghosts and antifields/antighosts and the BRST operator $Q$, hence we also have the formal representation

\begin{equation}\label{eq:FreeS2}
S_0={1\over 2}\langle\Psi|Q|\Psi\rangle|_{{\rm gh_m}=0}.
\end{equation}

This representation of $S$ can be used as a canonical generator of BRST-transformations according to equation (\ref{eq:CanonicalS}). A sample calculation suffices to bring this out. As an example, consider

\begin{equation}\label{eq:SampleCalculation1}
\delta_s|\Phi\rangle=(|\Phi\rangle,{1\over 2}\langle\Phi|Q|\Phi\rangle+\langle\Phi^\#|Q|{\cal C}\rangle)=
(|\Phi\rangle,\langle\Phi^\#|)Q|{\cal C}\rangle=Q|{\cal C}\rangle,
\end{equation}

where, according to (\ref{eq:FieldAntiFieldBracket}) and the correspondences of table (\ref{tab:Fields}), the only non-zero bracket is $(|\Phi\rangle,\langle\Phi^\#|)=1$. Note that the mechanics bra and ket structure plays no role here. The antibracket $(\cdot\,,\cdot)$ is completely indifferent to the Fock complex canonical structure of mechanical oscillators and ghosts. This also makes sense from the point of view that the abstract operation $^\#$ of forming antifield and antighosts has nothing to do with neither complex nor hermitean conjugation. Clearly, the rest of the BRST-transformations of (\ref{eq:BRSTTransformations1}) can be derived similarly.

To conclude, we have lifted the (mechanics BRST-based) {\it gauge invariant} $\langle\Phi|Q|\Phi\rangle$ free field theory into a (still mechanics BRST-based) {\it field theory BRST-invariant} $\langle\Psi|Q|\Psi\rangle$ free theory in the BV-formalism. The free field theory operator ${\bf s}$ is exactly represented by the mechanics operator $Q$. The possibility to maintain this precise correspondence lends strength to the formalism.

%*******************************
%MECHANICS BRST VS. BV FORMALISM
%*******************************************************
\subsection{The choice of formalism}\label{subseq: TCOF}
The choice of formalism may seem like a trivial point, but its importance should not be underestimated, as perhaps the fate of the Newtonian vs. the Leibnizian version of the calculus, no other comparisons implied, indeed shows. In the present context, we are actually laboring with three different choices of formal representation, and we will now take a moment to compare them. First of all, we will denote the gauge invariant action with the letter $A$, and in particular $A_0$ for the free theory. The field theory Batalin Vilkovisky BRST-invariant action will be denoted by $S$ and $S_0$ respectively. Thus $S_0^{(0)}=A_0$. 

\subsubsection*{DeWitt formalism}
Let us return to and begin with the DeWitt formalism as it is the most common in use in abstract discussions of field theory. Its main advantage is that it hides many of the complexities of field theory, by collecting all indices as well as spacetime dependence, in one abstract index. It is used in many treatises on general gauge theory, and it provides for compact formulas bringing out the essential points. However, it is not so convenient for practical calculations, because the abstract index must then be expanded into particular sets of indices.

\subsubsection*{Mechanics BRST formalism}
The mechanics BRST formalism utilizes the higher spin fiber of section \ref{subsec:ASHSF}, or perhaps a similar construction if a more elaborate fiber corresponding to other field multiplets is required. Now all indices, and spacetime dependence, is hidden in the object $|\Phi\rangle$. As compared to the DeWitt formalism it is more abstract (no indices), yet more concrete, since we actually have a concrete expansion in the Fock complex of the underlying mechanics model. This means, for instance, that the simple equation $Q|\Phi\rangle=0$ actually implies all of Fronsdal's wave equations \cite{Fronsdal1978}. Since $Q$ and $|\Phi\rangle$ have general properties that can be listed, we can work in an exact way with them without expanding, while if need arises, we can expand them out and get results for particular components. Therefore this formalism is very powerful for computational purposes. Needless to say, these computations, being algebraic in nature can be computerized. 

This very discussion now carries over to the full interacting theory and we can write out the action as a power series

\begin{equation}\label{eq:VertexActionA}
A={1\over 2}\langle\Phi|Q|\Phi\rangle+\sum_{n=3}^\infty g^{n-2}\langle\Phi|^{\otimes n}{\cal F}_n|-\rangle_{1\cdots n}.
\end{equation}

Everything here is well defined, except that for the vertex operators ${\cal F}_n$ we only have an ansatz, but it is a definite ansatz \cite{AKHB2005a} (see also section \ref{sec:VertexImplementation}). A similar expression can be written for the gauge transformations $\delta_\Xi|\Phi\rangle$. Demanding invariance of the action $\delta_\Xi A=0$ yields recursive equations for the unknown operators ${\cal F}_n$. What remains to be done here is to write the software to compute ${\cal F}_n$. Note that the power series here is formal, as is all power series expansions considered in this paper, by which we mean that questions of convergence are not considered.

This formalism is now easily lifted to field theory BRST-BV formalism, we just replace the fields $|\Phi\rangle$ with the fields$|\Psi\rangle$, denote the vertex operators with ${\cal V}_n$ and only keep terms at ghost number 0.

\begin{equation}\label{eq:VertexActionS1}
S={1\over 2}\langle\Psi|Q|\Psi\rangle|_{{\rm gh_m}=0}+\sum_{n=3}^\infty g^{n-2}\langle\Psi|^{\otimes n}{\cal V}_n|-\rangle_{1\cdots n}|_{{\rm gh_m}=0}.
\end{equation}

Gauge invariance of the action is now replaced by the master equation $(S,S)=0$. We will return to and elaborate this version of the theory below. The formula (\ref{eq:VertexActionS1}) contains more information than immediately meets the eye as will be apparent when it is expanded according to antighost number (see sections \ref{subsec:HPT} and \ref{subsec:HOVI}).

\subsubsection*{Abstract interface formalism}
In the abstract interface formalism, everything is hidden behind one symbol $\Phi$ that can be thought of as an interface, in the computer science meaning of the term, towards any implementation. Nothing prevents us from listing desired properties for the interface. Any implementation must then of course provide for these properties in terms of concrete data structures. We know that we are on firm ground, since the mechanics BRST formalism actually provides us with one well defined implementation. The strength of the abstract interface is that it frees us from any assumption about multiplet structure or spacetime geometry. 

One aspect of the theory that the abstract formalism clearly brings out is that in any non-terminating formal power series expansion of the interactions, the consecutive orders of the vertices fall into the structure of a strongly homotopy Lie algebra $L_\infty$. This result has been obtained before by more elaborate methods \cite{FulpLadaStasheff2002}. In the abstract formalism it is almost immediate, and certainly inevitable. Depending on the outlook adopted, it is either very deep, or just a superficial consequence of gauge invariance of formal power series.

In this formalism, focusing as it does on outward structure rather than internal machinery, the step to the language of category theory is very small, and we will briefly return to this topic in section \ref{sec:Categories}. 

However, I do not think that too much significance should be attached to any of these formalisms. In the theory of computation it is well known that any of a countable set of formalisms for effective computation all capture the same underlying concept of algorithmic computation. The same is probably true here. If interacting higher spin gauge field theory exists as a physical theory, any well tailored formalism, powerful enough to control the complexity, should be able to capture its properties and provide for effective calculation. It is just a matter of finding such a formalism. Of course, it could well be that a consistent theory do not exist, then the formalism should provide us with a no-go theorem. 

No formalism to be used by humans, captures all aspects of a certain type of objects. That would make the formalism top heavy and cumbersome and there are always connotations that are not formalized. These can often only be gathered by close study. For formalisms used by computers it is the other way around. All aspects must be captured in an exact way, but then the prize to pay is a certain rigidness, and only a limited number of properties can be implemented.

It is my impression that often to little care is exercised in setting up formalistic frameworks in the present context. Since all of the above described formalisms capture different aspects of higher spin gauge theory, we will switch between them as seems appropriate. 

\pagebreak
%===================================================================================================================
%DEFORMATION THEORY AND GAUGE INTERACTIONS
%===================================================================================================================
\section{Deformation theory and gauge interactions}\label{sec:DeformationTheory}
In this section we will try to organize the deformation theory of higher spin gauge fields to make it amenable to a systematic computation of the interaction vertices. We will start by looking in detail at the cubic and quartic vertices in order to gain intuition, and then treat all orders. The relation between vertex order perturbation theory and homological perturbation theory will be clarified. A formal proof of existence will be reviewed.  Before proceeding, some general words on the question of existence is perhaps appropriate.

%*****************************************
%DEFORMATION THEORY AND GAUGE INTERACTIONS
%***************************************************************
\subsection{A note on the existence question}\label{subsec:NEQ}
The existence of self-interacting higher spin gauge field theories in Minkowski background is neither proved nor disproved, if by proved we mean an explicit construction, and by disproved we mean an exhaustive search through all possible cases. It is simply still an open question, although the evidence from the investigations carried out during the last twenty years or so start to weigh in on the side of existence. Light-front cubic vertices exist for all integer spin. There is a covariant cubic vertex for spin 3, and new results for spin 3 and 4 cubic vertices are appearing (see \cite{BekaertBoulangerCnockaertLeclercq2006a} for references on recent work). There are well-defined mathematical structures governing power series expansions of the interactions (sh-Lie algebras, BRST-BV formalism, operads), as well the implementation in terms of an underlying mechanics BRST-Fock complex of the present paper (see also \cite{AKHB1988,AKHB2005a}).

Considering the cubic spin 3 vertex of BBvD, these authors pointed out \cite{Burgers1985thesis} that the non-closure of the pure spin 3 gauge algebra could in principle be remedied by introducing a spin 5 field. Then, if the pure spin 5 cubic interaction exists, then most likely its gauge algebra would not close unless spin 9 fields were introduced. It could then be conjectured that a spin 3 - spin 5 cubic interaction would require spin 7 fields to close that part of the gauge algebra. If a scenario like this were to be born out by explicit calculation, it could be conjectured that the structure so revealed would fall into the pattern of homological perturbation theory. The non-closure at the spin 3 gauge algebra level viewed as an obstruction is killed by introducing the spin 5 variation $\delta\varphi_5$. Of course, working explicitly like this without any supporting structure would be exceedingly impractical due to the rapidly increasing complication of the terms. This is one reason why we prefer to include all higher spin gauge fields from the outset, even though we still do not know a priori what multiplets of fields to choose. 

%*****************************************
%DEFORMATION THEORY AND GAUGE INTERACTIONS
%*************************************************************
\subsection{Deforming the Noether equation}\label{subsec:DNE}
Let us begin by writing down the relevant formulas in the standard notation as a backdrop to the BRST-Fock complex implementation. We are initially considering gauge transformations of the physical fields $\varphi^i$

\begin{equation}\label{eq:DeltaPhi}
\delta_\xi\varphi^i=R^i_\alpha\xi^\alpha,
\end{equation}

where all derivatives and all field dependence, possible to all orders, are packaged into $R^i_\alpha$. If we want to focus on a certain order, we decorate with subscripts ''$_n$'', taking ''$_0$'' to denote the free theory. The gauge parameter $\xi$ is of little importance since all algebraically interesting information is stored in the $R$'s. 

Invariance of the action $A$ is expressed by the Noether equation

\begin{equation}\label{eq:Noether1}
{\delta A\over \delta\varphi^i}\delta_\xi\varphi^i=0|_{\forall\xi}\quad\Rightarrow\quad{\delta A\over \delta\varphi^i}R^{i}_\alpha=0.
\end{equation}

Then expanding both the action 

\begin{equation}\label{eq:ExpandedAction1}
A=A_0+gA_1+g^2A_2+\ldots,
\end{equation}

and the gauge functionals

\begin{equation}\label{eq:ExpandedGaugeTransformations}
R^{i}_\alpha=R^i_{0\alpha}+gR^i_{1\alpha}+g^2R^i_{2\alpha}+\ldots,
\end{equation}

we get

\begin{equation}\label{eq:NoetherExpansion}
{\delta (A_0+gA_1+g^2A_2+\ldots)\over \delta\varphi^i}(R^i_{0\alpha}+gR^i_{1\alpha}+g^2R^i_{2\alpha}+\ldots)=0.
\end{equation}

Given a set of free fields, a free action and free gauge transformations, one could make assumptions (modulo field redefinitions) as to the form of the higher order terms and then try to solve equation (\ref{eq:NoetherExpansion}) order by order in $g$. There might be further consistency conditions to take into account, but the basic task remains the same. A detailed account of this approach can be found in \cite{Anco1992}, but rather than pursuing this further, we will turn at once to the BV formulation.

Starting from a generic free field theory action $A_0[\varphi^i]=S_0^{(0)}[\varphi^i]$, ghosts and antifields are introduced, and the action $S[\psi^A,\psi^\#_A]$ for the full theory is sought as a solution the master equation

\begin{equation}\label{eq:MasterEquation1}
(S,S)=0,
\end{equation}

starting out with

\begin{equation}\label{eq:FirstFewTermsAction}
S=S_0^{(0)}+\varphi^\#_iR^i_{0\alpha} C^\alpha + \ldots=S_0^{(0)} + S_0^{(1)}+ \ldots,
\end{equation}

where the dots stand for whatever comes next in the would be solution. The parentherized superscripts count the antighost number to disambiguate from the subscripts denoting interaction order.

Thus we see that the free theory action $S_0^{(0)}$ and the free theory gauge transformations $\delta_\xi\varphi^i=R^i_{0\alpha}\xi^\alpha$ together play the role of a base for a iterative, order by order, formal power series solution to the master equation (\ref{eq:MasterEquation1}). 

Now, two circumstances need to be kept in mind in order to avoid confusion when perusing the literature. First, finding a solution to the master equation amounts to lifting a gauge invariant theory into a BRST invariant theory in the space of fields and antifields. In that case we may already know the full gauge invariant theory, i.e. we do not really start from a {\it free} theory, but rather as an example, the full {\it interacting} Yang-Mills theory. This is the context in which the BRST-BV formalism was first developed. For reviews and full lists of references, see \cite{BarnichBrandtHenneaux2000,GomisParisSamuel}. Second, the expansion used in proving the existence of $S$ is an expansion in antighost number. However the expansion that we are really interested in when trying to deform a free theory into an interacting one, is the vertex order expansion, i.e. an expansion in the number of fields. Thus, in order to use the BV-formalism as a framework for deriving interactions by deforming a free field theory, we must understand how these two expansions correlate. As a first step we will look in some detail at the cubic and quartic level.

%*****************************************
%DEFORMATION THEORY AND GAUGE INTERACTIONS
%*****************************************************************************
\subsection{Power counting in the vertex implementation}\label{subsec:PCITVI}
In order to better understand the formal power series expansion of S, we will express the $n$-th point interaction in terms of the vertex implementation \cite{AKHB2005a}. Thus, as in (\ref{eq:VertexActionA}), we write the $n$-point vertex as 

\begin{equation}\label{eq:VertexShorthand}
g^{n-2}\langle\Phi|^{\otimes n}{\cal F}_n|-\rangle_{1\cdots n},
\end{equation}

where ${\cal F}_n$ is an operator to be determined in terms of the Fock complex creation operators and momenta. Its detailed form will be discussed in section \ref{sec:VertexImplementation}. For now it is enough to know that is has mechanics ghost number ${\rm gh_m}=0$ and has mass dimensionality $d=4-n$ in $D=4$. The expression $|-\rangle_{1\cdots n}$ is short for

\begin{equation}\label{eq:VertexVacuum}
|-\rangle_{1\cdots n}=\prod_{i=1}^n\int d^4p_i|-\rangle_i \delta^4\Big(\Sigma_{i=1}^n p_i\Big)\quad{\stackrel{Fourier}{\simeq}}\quad\int d^4x\prod_{i=1}^n|-\rangle_i.
\end{equation}

We will generally just write $|{\cal F}_n\rangle$ as a shorthand for ${\cal F}_n|-\rangle_{1\cdots n}$, or as $|{\cal F}_{1\cdots n}\rangle$ when we want to explicitly focus on the dummy indices $1,2,\cdots,n$.

We want to derive general constraints on the numbers of different fields in the $n$-point interaction term. To that end, formally replace $\langle\Phi|$ in (\ref{eq:VertexShorthand}) with $\langle\Psi|$ according to (\ref{eq:FormalPsi}), simplifying the notation

\begin{equation}\label{eq:SimpleNotationPhiMany}
\langle\Psi|^n=\langle\Phi|^a\langle\Phi^\#|^b\langle{\cal C}|^c\langle{\cal C}^\#|^d.
\end{equation}

Thus consider the somewhat schematic form for $S_{n-2}$ 

\begin{equation}\label{eq:SN}
g^{n-2}\langle\Phi|^a\langle\Phi^\#|^b\langle{\cal C}|^c\langle{\cal C}^\#|^d|{\cal V}_n\rangle,
\end{equation}

where we have replaced the vertex operator ${\cal F}_n$ with ${\cal V}_n$. Strictly speaking this is not necessary, since what we have for ${\cal F}_n$ is the most general ansatz, thus as far as that goes, nothing more can be done. However, we want to acknowledge the possibility that the actual solutions to ${\cal F}_n$ and ${\cal V}_n$ might differ.

Counting number of fields, ghost number and mass dimensionality, we arrive at the following equations governing the field content

\begin{eqnarray}
a+b+c+d&=&n\quad\mbox{Field count}\label{eq:FCE1}\\
a-b+3c-3d&=&n\quad\mbox{Mechanics ghost number count}\label{eq:FCE2}\\
a+3b-c+5d&=&n\quad\mbox{Mass dimensionality count}\label{eq:FCE3}
\end{eqnarray}

These equations are not independent, as is obvious upon adding the last two. Counting field theory ghost number is equivalent to counting mechanics ghost number. Furthermore, subtracting the two last equations, yields $c-b=2d$. Thus, $c-b$  and consequently $b+c$ is an even number, ensuring that $S_n$ is Grassman even. The total antighost number is equal to $b+2d$ which works out to simply $c$.

%*****************************************
%DEFORMATION THEORY AND GAUGE INTERACTIONS
%*****************************************************
\subsection{Vertex order expansion}\label{subsec:VOE}
In the BV formalism, both the action and the BRST transformations are unified in the generator $S[\psi,\psi^\#]$, and the deformed Noether equations (\ref{eq:Noether1}) - (\ref{eq:NoetherExpansion}) can be more compactly expressed in terms of a deformation of $S$

\begin{equation}\label{eq:DeformationOfS}
S=S_0+gS_1+g^2S_2+\ldots,
\end{equation}

where $S_n$ is homogenous of degree $n+2$ in all the fields $\psi,\psi^\#$. $S$ should satisfy the master equation (\ref{eq:MasterEquation1}) to all orders in $g$

\begin{equation}\label{eq:MasterEquation2}
(S_0+gS_1+\ldots,S_0+gS_1+\ldots)=0.
\end{equation}

Spelling this out for the first few orders in $g$ yields

\begin{eqnarray}\label{eq:S0S0}
(S_0,S_0)=0, \label{eq:S0S0} \\
(S_0,S_1)=0, \label{eq:S0S1} \\
(S_1,S_1)+2(S_0,S_2)=0, \label{eq:S1S1} \\
(S_1,S_2)+(S_0,S_3)=0, \label{eq:S1S2} 
\end{eqnarray}

the general equation being

\begin{equation}\label{eq:S0SN}
(S_0,S_n)+{1\over 2}\sum_{k=1}^{n-1}(S_{k},S_{n-k})=0\quad\mbox{ for }n=0,1,2,\ldots.
\end{equation}

The outward appearance of these equations is clearly that of homology theory. In section \ref{subsec:FNOP} we will reproduce the proof that they can be solved in formal perturbation theory. 

Eventually we want to set all the antifields to zero and retrieve the gauge invariant interacting field theory

\begin{equation}\label{eq:SetPhiHashZero}
A[\varphi]=S^{(0)}[\varphi]=S[\psi,\psi^\#]|_{\psi^\#=0}.
\end{equation}

%*****************************************
%DEFORMATION THEORY AND GAUGE INTERACTIONS
%*********************************************************************************
\subsection{Understanding the cubic and quartic interaction}\label{subsec:UTFOII}
To be concrete and see how the formalism works in practice, let us take a closer look at the form of the cubic and quartic vertices in this formulation. From the power counting formulas we get the possible combinations of fields as recorded in tables \ref{tab:CubicVertexNumbers} and \ref{tab:4and5VertexNumbers}.

\vskip0.2cm
\begin{table}[h]\label{tab:CubicVertexNumbers}
\begin{tabular}{lllll}a&b&c&d&Fields  \\ \hline
3&0&0&0&$|\Phi\rangle|\Phi\rangle|\Phi\rangle$ \\
1&1&1&0& $|\Phi\rangle|\Phi^\#\rangle|{\cal C}\rangle$ \\
0&0&2&1& $|{\cal C}\rangle|{\cal C}\rangle|{\cal C}^\#\rangle$
\end{tabular}
\caption{Field content in $S_1$}
\label{tab:CubicVertexNumbers}
\end{table}

\vskip0.2cm 
\begin{table}[h]
\begin{tabular}{lllll}
a&b&c&d&Fields  \\ \hline
4&0&0&0& $|\Phi\rangle|\Phi\rangle|\Phi\rangle|\Phi\rangle$ \\
2&1&1&0& $|\Phi\rangle|\Phi\rangle|\Phi^\#\rangle|{\cal C}\rangle$ \\
1&0&2&1& $|\Phi\rangle|{\cal C}\rangle|{\cal C}\rangle|{\cal C}^\#\rangle$ \\
0&2&2&0& $|\Phi^\#\rangle|\Phi^\#\rangle|{\cal C}\rangle|{\cal C}\rangle$
\end{tabular}
\caption{Field content in $S_2$}
\label{tab:4and5VertexNumbers}
\end{table}
\vskip0.2cm 

The fields entering the interaction terms are labeled by integers $1,2,3,\ldots$, or rather, their momenta and Fock complexes spanning the respective fibers are labeled by these numbers. This numbering has no intrinsic meaning of course, and the vertices should be fully symmetric under permutations of the labels. An easy way to achieve this is to simply sum each term over all {\it different} permutations with weight 1. The weighting is a question of convenience, another choice could be to weigh by ${1\over n!}$, a choice that would lead to a simpler end result, but which is inconvenient to carry around in actual computations (see the discussion in section \ref{subsec:HOVI} where we will actually rescale the vertices with factorials). A few examples should suffice to bring out the principle. Thus, in the cubic vertex there is just one term $\langle\Phi_1|\Phi_2|\langle\Phi_3||{\cal V}_{123}\rangle$, since all other permutations can be reached by rearranging the fields. In the quartic vertex, the combination of fields $|\Phi^\#\rangle|\Phi^\#\rangle|{\cal C}\rangle|{\cal C}\rangle$ contributes to the vertex with six terms different by permutations

\begin{eqnarray}\label{eq:ExampleFieldPerm}
\langle\Phi^\#_1|\langle\Phi^\#_2|\langle{\cal C}_3|\langle{\cal C}_4||{\cal V}_{1234}\rangle&+&\nonumber
\langle\Phi^\#_1|\langle\Phi^\#_3|\langle{\cal C}_2|\langle{\cal C}_4||{\cal V}_{1234}\rangle+\\\nonumber
\langle\Phi^\#_1|\langle\Phi^\#_4|\langle{\cal C}_2|\langle{\cal C}_3||{\cal V}_{1234}\rangle&+&
\langle\Phi^\#_2|\langle\Phi^\#_3|\langle{\cal C}_1|\langle{\cal C}_4||{\cal V}_{1234}\rangle+\\
\langle\Phi^\#_2|\langle\Phi^\#_4|\langle{\cal C}_1|\langle{\cal C}_3||{\cal V}_{1234}\rangle&+&
\langle\Phi^\#_3|\langle\Phi^\#_4|\langle{\cal C}_1|\langle{\cal C}_2||{\cal V}_{1234}\rangle.
\end{eqnarray}

The rest of the $4!$ permutations can be reached by rearranging the fields. The question of Grassman signs does not arise due to the nice conspiration of the Grassman properties of the fields/antifields themselves and the first quantized Fock complex ghosts which effectively makes all the Fock complex fibers $|\Phi\rangle,|\Phi^\#\rangle,|{\cal C}\rangle$ and $|{\cal C}^\#\rangle$ even (see comment below table \ref{tab:Fields}). We will simply denote this by a summation symbol 

\begin{equation}\label{eq:SumSymbol}
\sum_{\rm{dp}(i_1\cdots i_n)}
\end{equation}

where ${i_1\cdots  i_n}$ symbolizes a set of $n$ different indices, and then decorate each type of term with a superscript $[P]$ denoting the number of different permutations in the sum. When $P=1$ we may drop the summation symbol and the superscript. 

Thus, up to quartic order, we have 

\begin{eqnarray}\label{eq:StoCubicOrder}
S={1\over 2}\langle\Phi|Q|\Phi\rangle + \langle\Phi^\#|Q|{\cal C}\rangle +  \nonumber\\
g\sum_{\rm{dp}(123)}\Big(\langle\Phi_1|\langle\Phi_2|\langle\Phi_3|{\cal V}_{123}\rangle +
\langle\Phi_1|\langle\Phi^\#_2|\langle{\cal C}_3|{\cal V}_{123}\rangle^{[6]} +
\langle{\cal C}_1|\langle{\cal C}_2|\langle{\cal C}^\#_3|{\cal V}_{123}\rangle^{[3]}\Big)+ \nonumber\\
g^2\sum_{\rm{dp}(1234)}\Big(\langle\Phi_1|\langle\Phi_2|\langle\Phi_3|\langle\Phi_4|{\cal V}_{1234}\rangle+
\langle\Phi_1|\langle\Phi_2|\langle\Phi^\#_3|\langle{\cal C}_4|{\cal V}_{1234}\rangle^{[12]}- \nonumber\\
\langle\Phi_1|\langle{\cal C}_2|\langle{\cal C}_3|\langle{\cal C}^\#_4|{\cal V}_{1234}\rangle^{[12]}+
\langle\Phi^\#_1|\langle\Phi^\#_2|\langle{\cal C}_3|\langle{\cal C}_4|{\cal V}_{1234}\rangle^{[6]}\Big).\nonumber
\end{eqnarray}
\begin{equation}\end{equation}

The reason that we can use the same vertex operators in all three cubic terms and all four quartic terms is the very generality of ${\cal V}_3$ and ${\cal V}_4$ and the fact that, corresponding to the Fock complex expansions of the fields, they will actually pick out different parts of the vertices $|{\cal V}\rangle$.

We can now compute $(S,S)$ to first and second order in the coupling $g$, using 

\begin{equation}\label{eq:SSG1}
(\Phi_i,\Phi^\#_j)=\delta_{ij},\quad({\cal C}_i,{\cal C}^\#_j)=\delta_{ij},
\end{equation}

where, as already noted, the bra-ket structure plays no role. Indeed, since all terms in the expansion of $S$ are real, we can always conjugate the terms entering $(S,S)$ so as to bracket off a bra with a ket, keeping the balance of total number of bras and kets intact. In practice, this never leads to any problems, and we chose not to introduce formalism to keep track of it. 

Corresponding to equation (\ref{eq:S0S1}) we get, modulo permutations of the index numbers

\begin{eqnarray}\label{eq:}
\langle\Phi_1|\langle\Phi_2|\langle{\cal C}_3|\sum_{r=1}^3Q_r|{\cal V}_{123}\rangle=0,\\
\langle\Phi_1^\#|\langle{\cal C}_2|\langle{\cal C}_3|\sum_{r=1}^3Q_r|{\cal V}_{123}\rangle=0.
\end{eqnarray}

Thus, BRST-invariance to $g$ order amounts to 

\begin{equation}\label{eq:V3Eqs}
\sum_{r=1}^3 Q_r|{\cal V}_{3}\rangle=0,
\end{equation}

which is the same equation as follows from gauge invariance. It is known to reproduce cubic order Yang-Mills  \cite{AKHB1988} given a few more considerations that we will return to in section \ref{subsec:GSASC}. 

Next, corresponding to equation (\ref{eq:S1S1}) we get to order $g^2$, modulo permutations of the index numbers

\begin{eqnarray}
8\langle\Phi_1|\langle\Phi_2|\langle\Phi_3|\langle{\cal C}_4|\sum_{r=1}^4 Q_r|{\cal V}_{1234}\rangle=36\langle\Phi_1|\langle\Phi_2|\langle\Phi_3|\langle{\cal C}_4||{\cal V}_{12x}\rangle|{\cal V}_{34x}\rangle,\label{eq:PPPC}\\
24\langle\Phi_1|\langle\Phi^\#_2|\langle{\cal C}_3|\langle{\cal C}_4|\sum_{r=1}^4 Q_r|{\cal V}_{1234}\rangle=108\langle\Phi_1|\langle\Phi^\#_2|\langle{\cal C}_3|\langle{\cal C}_4||{\cal V}_{12x}\rangle|{\cal V}_{34x}\rangle,\\
8\langle{\cal C}_1|\langle{\cal C}_2|\langle{\cal C}^\#_3|\langle{\cal C}^\#_4|\sum_{r=1}^4 Q_r|{\cal V}_{1234}\rangle=36\langle{\cal C}_1|\langle{\cal C}_2|\langle{\cal C}^\#_3|\langle{\cal C}^\#_4||{\cal V}_{12x}\rangle|{\cal V}_{34x}\rangle,
\end{eqnarray}

where we have recorded numerical factors resulting from the combinatorics as they actually appear in the computation. These three equations are all implied by

\begin{equation}\label{eq:V4Eqs}
\sum_{r=1}^4 Q_r|{\cal V}_{1234}\rangle=-{9\over 2}\Big[{4\choose 2}{1\over 2}\Big]^{-1}|{\cal V}_{12x}\rangle\diamond|{\cal V}_{34x}\rangle
\end{equation}

with an extra combinatorial factor and $\diamond$ denoting a symmetrized Fock space contraction. The notation is a bit inconsequential. When we write $|{\cal V}_{12x}\rangle|{\cal V}_{34x}\rangle$, we imply an inner product contraction over the Fock space numbered by $x$. This means that the expression $\langle\Phi_1|\langle\Phi_2|\langle\Phi_3|\langle{\cal C}_4||{\cal V}_{12x}\rangle|{\cal V}_{34x}\rangle$ in equation (\ref{eq:PPPC}) should really be written as $\langle{\cal V}_{12x}|\Phi_1\rangle|\Phi_2\rangle\langle\Phi_3|\langle{\cal C}_4||{\cal V}_{34x}\rangle$. Let us simply introduce the convention that when an index $x$ appears twice in a combination $|{\cal V}_{i_1\cdots x\cdots i_n}\rangle|{\cal V}_{j_1\cdots x\cdots j_m}\rangle$, then one of the Fock spaces should be switched ${\rm bra}\leftrightarrow{\rm ket}$ upon which the contractions can be performed. In a similar way, $|{\cal V}_{i_1\cdots x\cdots i_n}\rangle\diamond|{\cal V}_{j_1\cdots x\cdots j_m}\rangle$ denotes a symmetrized Fock space contraction over index $x$, implicitly containing ${n+m-2\choose n-1}(1-{1\over 2}\delta_{nm})$ terms in order to achieve full permutational symmetry. This explains the extra numerical factor in equation (\ref{eq:V4Eqs}).

Sometimes it is convenient use a dot to indicate a contraction of a pair of Fock spaces between two vertices $|{\cal V}_n\rangle\cdot|{\cal V}_m\rangle$. Likewise we could write a symmetrized $|{\cal V}_n\rangle\diamond|{\cal V}_m\rangle$. Due to the permutational symmetry there are no ambiguities here.

The formulas for the cubic (\ref{eq:V3Eqs}) and quartic vertices (\ref{eq:V4Eqs}) derived here will be generalized to all orders in section \ref{subsec:HOVI}. It is apparent from the calculations shown here that a vertex order expansion mixes terms with different antighost numbers. This we will now look deeper into.

%*****************************************
%DEFORMATION THEORY AND GAUGE INTERACTIONS
%**************************************************************
\subsection{Homological perturbation theory}\label{subsec:HPT}
In order to use homological perturbation theory and apply the existence theorem for the gauge algebra to show that $S$ formally exists to all orders in the interaction, we must understand how the vertex order expansion of $S$ relates to the antighost number expansion. This will greatly clarify the structure of the theory. To that end, consider the expansion in antighost number

\begin{equation}\label{eq:AntiGhostNumberExpansionS}
S=S^{(0)}+S^{(1)}+\ldots=\sum_{k\geq0}S^{(k)}.
\end{equation}

Since the antighost number is actually equal to the number of pure ghost fields in the terms, this is really an expansion in ${\cal C}$. Temporarily reverting to DeWitt notation in terms of generic fields $\varphi$ and ghosts $C$, we have

\begin{eqnarray}\label{eq:StandardExpansionS}
S=S^{(0)}(\varphi)+C^\alpha R^i_\alpha(\varphi)\varphi^\#_i&+& \nonumber\\
C^\alpha C^\beta T_{\alpha\beta}^\gamma(\varphi)C^\#_\gamma+C^\alpha C^\beta M_{\alpha\beta}^{ij}(\varphi)\varphi^\#_i\varphi^\#_j&+&{\cal O}(C^3).
\end{eqnarray}

For simplicity we just consider the irreducible case here. In this para-meterization of $S$, $S^{(0)}(\varphi)$ denotes the action to all orders of interaction. The next term is $S^{(1)}=C^\alpha R^i_\alpha(\varphi)\varphi^\#_i$ where $R^i_\alpha(\varphi)$ denotes the gauge transformations to all orders in $g$. Next, $T_{\alpha\beta}^\gamma(\varphi)$ are the structure functions 

\begin{equation}\label{eq:StandardGaugeAlgebra}
R^j_\alpha{\delta R^i_\beta\over\delta\varphi^j}-R^j_\beta{\delta R^i_\alpha\over\delta\varphi^j}=T^\gamma_{\alpha\beta}R^i_\gamma+\ldots,
\end{equation}

where the dots indicate that in general there are trivial gauge transformations that vanish on the stationary surface (the open gauge algebra case). In the reducible case, obstructions would occur, and the procedure would then be to add new fields, the variations of which, properly defined, would kill the obstructions \cite{Tate1956}. 

Equation (\ref{eq:StandardGaugeAlgebra}) as well as the Noether identities (\ref{eq:Noether1}) and the higher order structure equations of the gauge algebra are all generated by the single nil-potency condition $(S,S)=0$ on $S$. Of course, nothing is for free, $(S,S)$ does not vanish {\it unless} the structure functions $T^\gamma_{\alpha\beta},M_{\alpha\beta}^{ij}$, et cetera, all exist and satisfy the structure equations {\it precisely} parameterized by the master equation. So given a consistent gauge invariant interacting field theory, all its gauge structure can be nicely collected in the cohomology of a single differential $S$. This could be phrased as saying that the field/antifield BV structure is built {\it on the outside} of an already consistent gauge theory. It relies on the form of the action and gauge transformations, but it does not affect them in any way, but just organizes the consistency requirements in an efficient way. 

This is where the existence proof of Batalin and Vilkovisky for the gauge algebra enters. Given a few general conditions to be met by general gauge theories, the solutions to the equations for the structure functions generated by the master equation can actually be shown to exist \cite{BatalinVilkovisky1985a,FischHenneaux1990a,FischHenneauxStasheffTeitelboim1989a}. These general conditions, called ''boundary conditions'', amounts to $S^{(0)}=A$ and $S^{(1)}=\phi_i^{\#}R^i_\alpha C^\alpha$, i.e. the lowest orders in the antighost number expansion should correspond to the classical action $A$ and the gauge transformations $R$. The existence theorem says nothing about the {\it existence of interactions}, but of course, since for a free theory it is rather trivial, its importance lies in the existence of the higher order structure functions of the gauge algebra when interactions are present. 

Now in the case at hand, we don't really have an interacting theory. What we have is a power series expansion of the interactions with as yet undetermined coefficients. The existence theorem then guarantees the existence of power series expansions of the higher order (corresponding to antighost number) structure functions. Thus, given the existence proof, it is only a matter of calculation to recursively compute successive powers $S_n$ of $S$ from the equations (\ref{eq:S0SN}). Even without an existence proof, the calculation can be attempted, and if there is a solution it will be found, at least up to a certain power, including the higher order (antighost number) structure functions. But as far as we can see, the existence theorem, or rather its standard proof, does not guarantee the existence of interactions without further ado. All is not lost though, as shown in \cite{GomisParis1993} the existence proof can be carried through in the vertex order expansion. The argument will be reviewed in section \ref{subsec:FNOP}.

Let us now organize the expansion of the master equation according to antighost number. To that end, note that while $(\Phi,\Phi^\#)$ decreases antighost number by 1, $({\cal C},{\cal C}^\#)$ decreases antighost number by 2. Thus $(S^{(0)},S^{(n)})$ decreases antighost number by 1 since $S^{(0)}$ only contains fields $\Phi$. Using the notation

\begin{equation}\label{eq:RN}
R^{(n-1)}=\sum_{k=1}^{n-1}S^{(k)}
\end{equation}

we get

\begin{equation}\label{eq:S0SNAgh}
2(S^{(0)},S^{(n)})+(D^{(n-1)},D^{(n-1)})|_{({\rm gh}_a=n-1)}=0\quad\mbox{ for }n=1,2,\ldots,
\end{equation}

the first two cases of which are explicitly

\begin{eqnarray}\label{eq:S0S0Agh}
(S^{(0)},S^{(1)})=0,\quad {\rm gh_a}=0, \label{eq:S0S1Agh} \\
2(S^{(0)},S^{(2)})+(S^{(1)},S^{(1)})+(S^{(1)},S^{(2)})=0,\quad {\rm gh_a}=1. \label{eq:S1S1Agh}
\end{eqnarray}

The equation $(S^{(0)},S^{(0)})=0$ is missing from the list as it is void of content. The first equation, at antighost number 0, contains much more information. Since, as noted above, $S^{(0)}$ encodes interactions to all orders and $S^{(1)}$ encodes the gauge transformations to all orders, this equation expresses gauge invariance to all orders. The second equation expresses consistency of the gauge algebra.

\subsubsection*{Relating antighost number expansion to vertex order expansion}\label{subsec:RANEVOP}
Based on the power counting formulas (\ref{eq:FCE1} - \ref{eq:FCE3}) we can now understand the relation between antighost number expansion and vertex order expansion. For a certain antighost number ${\rm gh_a}$ there is no upper limit to the vertex order $n$. Thus it makes sense to expand the classical action (${\rm gh}_a=0$), the gauge transformations (${\rm gh_a}=1$) and the higher order structure functions (${\rm gh_a}\geq 2$) in homogenous powers $n$ of all the fields and antifields at any order of ${\rm gh_a}$. There is however a lower limit to $n$, for instance for ${\rm gh_a}=2$, the vertex order expansion starts with the cubic term ${\cal C}^\#{\cal C}{\cal C}$. Conversely, for a given vertex order $n$, there is a maximum antighost number ${\rm gh_a}$ given by the integer part of $2n/3$.

%*****************************************
%DEFORMATION THEORY AND GAUGE INTERACTIONS
%*******************************************
\subsection{Higher orders in the vertex implementation}\label{subsec:HOVI}
With the confidence and intuition gained by understanding the relation between the antighost number and vertex order expansions, let now us refocus our discussion of $S$ and write out explicitly the first few terms in the antighost number expansion (\ref{eq:AntiGhostNumberExpansionS})

\begin{eqnarray}
S^{(0)}&=&{1\over 2}\langle\Phi|Q|\Phi\rangle+\sum_{n=3}^{\infty} g^{n-2}\langle\Phi|^{\otimes n}|{\cal V}_n\rangle,\\
S^{(1)}&=&\langle\Phi^\#|Q|{\cal C}\rangle+\sum_{n=3}^\infty g^{n-2}\langle\Phi^{\otimes(n-2)}\Phi^\#{\cal C}||{\cal V}_n\rangle^{[n(n-1)]},\\
S^{(2)}&=&\sum_{{\rm d.p}({n=3})}^{\infty} g^{n-2}\langle\Phi^{\otimes n-3}{\cal C}^\#{\cal C}{\cal C}||{\cal V}_n\rangle^{[n(n-1)(n-2)/2]}+\\
&+&\sum_{{\rm d.p}({n=4})}^{\infty} g^{n-2}\langle\Phi^{\otimes n-4}\Phi^\#\Phi^\#{\cal C}{\cal C}||{\cal V}_n\rangle^{[n(n-1)(n-2)(n-3)/4]}.
\end{eqnarray}

Clearly, by comparing to the generic form of $S$ (\ref{eq:StandardExpansionS}), $S^{(0)}$ is the action to all orders, i.e. the generator of the field equations. $S^{(1)}$ is the generator of gauge transformations to all orders, and $S^{(2)}$ generates the gauge algebra structure functions. We now see that all this structure is economically encoded in the vertex operators $|{\cal V}_n\rangle$. The antifields and ghost/antighost fields simply pick out different parts of the same object $|{\cal V}_n\rangle$. 

Returning next to the equations of homological perturbation theory of section \ref{subsec:HPT} we can now calculate equation (\ref{eq:S0S1Agh}), i.e. the master equation at ${\rm gh_a}=0$, with the result that at arbitrary vertex order $g^{(n-2)}$ we get

\begin{eqnarray}\label{eq:SumQonVertexField}
\langle\Phi^{(n-1)}{\cal C}|\sum_{r=1}^nQ_r|{\cal V}_n\rangle=\nonumber \\-\sum_{k=3}^{n-1}k(n+2-k)\langle\Phi^{(k-1)}|{\cal V}_k\rangle^{[1]}\cdot\langle\Phi^{(n-k)}{\cal C}|{\cal V}_{(n+2-k)}\rangle^{[n+1-k]}.
\end{eqnarray}

This equation, derived at antighost level zero, expresses gauge invariance of the action. Clearly the overall combination of fields $|\Phi_1\cdots\Phi_{n-1}{\cal C}_n\rangle$ and various permutations thereof, play no other role than keeping the equation pemutationally correct. Thus we would like to remove the fields while keeping track of permutations. This can be done by writing out the result in terms of the fully symmetrized $\diamond$-contractions between vertices defined in section \ref{subsec:UTFOII}.

Thus removing the fields, rearranging the terms and symmetrizing accordingly, yields

\begin{equation}\label{eq:SumQonVertex}
\sum_{r=1}^{n+1}Q_r|{\cal V}_{n+1}\rangle=-\sum_{p=0}^{\lfloor(n-3)/2\rfloor}C_{np}|{\cal V}_{p+3}\rangle\diamond|{\cal V}_{n-p}\rangle,
\end{equation}

where the combinatorial factor is 

\begin{equation}\label{eq:CombFactor}
C_{np}=\frac{(p+3)(n-p)}{{n+1\choose p+2}\big(1-\frac{1}{2}\delta_{(p+3)(n-p)}\big)}=\frac{(p+3)!(n-p)!}{(n+1)!\big(1-\frac{1}{2}\delta_{(p+3)(n-p)}\big)}.
\end{equation}

Now we can simplify by absorbing the factorials into the vertices according to $n!|{\cal V}_n\rangle\rightarrow|{\cal V}_n\rangle$, while the extra factor ${1\over 2}$, needed when the two vertices have the same number of indices, is implicit   in the $\diamond$-contraction. With this rescaling of the vertices, we get simply

\begin{equation}\label{eq:SumQonNewVertex}
\sum_{r=1}^{n+1}Q_r|{\cal V}_{n+1}\rangle=-\sum_{p=0}^{\lfloor(n-3)/2\rfloor}|{\cal V}_{p+3}\rangle\diamond|{\cal V}_{n-p}\rangle.
\end{equation}

This equation (or the non-rescaled (\ref{eq:SumQonVertex})) is however much stronger than (\ref{eq:SumQonVertexField}) as the fields only pick out parts of the vertices. It turns out that this equation implies $(S,S)$ to all orders in antighost number expansion. The reason is that at any antighost number level we just get the same equation (\ref{eq:SumQonVertex}) contracted with various combinations of fields. This is born out by calculating the next level of antighost number in the master equation, equation (\ref{eq:S1S1Agh}). Borrowing terminology from computer science, we could say that the vertices $|{\cal V}_n\rangle$ are {\it polymorphic}. This is more than mere analogy, as all interaction data to all orders is in fact encoded in the vertices $|{\cal V}_n\rangle$.

Thus, practical calculation shows that whether we calculate in the vertex order expansion or in the antighost number expansion, we get the same equations in the Fock complex vertex implementation. This is perhaps not so surprising after all. Equations (\ref{eq:SumQonNewVertex}) for $m\geq 3$, together with the equation (\ref{eq:V3Eqs}) for $m=2$, provide for the purely algebraic underpinning of higher spin gauge interactions. 

The next step is to show that these equations follow from the product identities of the abstract $L_\infty$ algebra. 

%*****************************************
%DEFORMATION THEORY AND GAUGE INTERACTIONS
%**************************************************************
\subsection{Semantic mapping of the sh-Lie structure}\label{subseq:EshL}
We will now adopt the point of view that the abstract interface formalism provides a syntax for higher spin gauge field theory, while the implementation in terms of Fock complex fibers and vertices (to be further developed in the next section) provides a concrete semantics. 

Phrased differently, we want to show that the sh-Lie structure is encoded in equations (\ref{eq:V3Eqs}) and (\ref{eq:SumQonNewVertex}). In order to do that, we will provide a semantic map from the abstract types $\Phi ::{\cal H}$ and ${\bf pr}::{\cal H}^\otimes\rightarrow{\cal H}$ to the concrete implementation in terms of $|\Phi\rangle$ and $|{\cal V}_n\rangle$. Remember that in the abstract interface, the multilinear and polymorphic product maps ${\bf pr}$ encode the interactions between the abstract fields.

Let us first write down the product identities of the sh-Lie algebra as they were derived in \cite{AKHB2005a} in the case of even abstract fields $\Phi$ (taking odd fields into account will not be required)

\begin{equation}\label{eq:ProductIdentity3}
\sum_{{k=0,l=0}\atop \chi(k,l)}^{k+l=n}{\bf pr}(\Phi_{i_1}\cdots\Phi_{i_k},{\bf pr}(\Phi_{j_1}\cdots\Phi_{j_l}))=0.
\end{equation}

Here, the index set $\{1,\cdots,n\}$ is split into the two sets $\{i_1,\cdots, i_k\}$ and $\{j_1,\cdots, j_l\}$. A particular split $\{\{i_1,\cdots, i_k\},\{j_1,\cdots, j_l\}\}$ is denoted by $\chi(k,l)$. The sum then runs over all different such splits. Note that we really mean sets here, not lists, as the order of the indices does not matter.

Even simpler, taking into account that the product maps are fully symmetric in all arguments, we can write the product identities as

\begin{equation}\label{eq:ProductIdentity2}
\sum_{{k=0,l=0}\atop \rm{cycl. perm.}}^{k+l=n}{\bf pr}(\Phi^k,{\bf pr}(\Phi^l))=0.
\end{equation}

Note that these sh-Lie identities were derived using only the simplest equational reasoning, employing nothing more advanced than substitutions and rearrangements of terms. Our goal is now to map them to the main recursive identities (\ref{eq:V3Eqs}, \ref{eq:SumQonNewVertex}) governing the vertices in the Fock complex implementation.

The product of $n$ fields evaluates to an object in the algebra, which is itself a field 

\begin{equation}\label{eq:ProductMap}
{\bf pr}(\Phi_1,\ldots,\Phi_n)\rightarrow\Phi_{n+1}.
\end{equation}

It will be convenient to introduce a special provision to deal with the one-product ${\bf pr}(\Phi)$ which in the abstract interface is naturally interpreted as a linear transformation $K\Phi$. Straining the formalism a little, the one-product can be made to conform to (\ref{eq:ProductMap}), if we write

\begin{equation}
K_2\Phi_2=K_2\Phi_1\delta_{12}={\bf pr}(\Phi_1)\rightarrow\Phi_2
\end{equation}

by which we mean that in whatever way we compute the one-product (or linear transformation), we change index on route. This is clearly not a natural thing to do, but it is helpful as will become apparent when we do the same in the Fock complex

\begin{equation}
Q_2|\Phi_2\rangle=Q_1|\Phi_2\rangle\delta_{12}=\langle\Phi_1|{\cal V}_2\rangle.
\end{equation}

In this way we can think of the action of the BRST operator $Q$ in terms of a 2-vertex $|{\cal V}_2\rangle$.

The semantic map can now be defined. The abstract fields $\Phi$ are simply mapped to Fock complex fibers $|\Phi\rangle$, and the products ${\bf pr}_n$ are mapped to vertices $|{\cal V}_{n+1}\rangle$. Indeed using an arrow $\hookrightarrow$ to denote the semantic map, we have

\begin{eqnarray}
\Phi_k\hookrightarrow|\Phi_k\rangle \quad \mbox{ for } k\in N\\
{\bf pr}(\Phi_1,\ldots,\Phi_n)\hookrightarrow\langle\Phi_1|\ldots\langle\Phi_n|{\cal V}_{n+1}\rangle \quad \mbox{ for } n\geq 1 .
\end{eqnarray}

Note that the last equation precisely utilizes our special provisions for the one-product, i.e. the case $n=1$. Note also that we are bit lenient with the formalism here, as the $(n+1)$-th Fock space in the vertex $|{\cal V}_{n+1}\rangle$ should really be switched to a bra. As it stands, the last equation produces a ket. It could be fixed at the cost of a more cumbersome notation, essentially by extending the semantic map to cover the abstract interface inner product ${\bf in}$. Since no problems arise in practice, we leave out this detail.

All this can be lifted from mechanics BRST to field theory BRST-BV by simply replacing $\Phi$ everywhere by $\Psi$. As pointed out above, the vertices are truly polymorphic, providing for the service of computing the appropriate product depending on the collection of fields $({\cal C},{\cal C}^\#,\Phi,\Phi^\#)$ contained in $\Psi$. Since the abstract fields are Grassman graded, we would have to use the generalized product identities taking Grassman sign factors into account. However, as already pointed out, the Fock complex fibers $(|{\cal C}\rangle,{|\cal C}^\#\rangle,|\Phi\rangle,|\Phi^\#\rangle)$ are all even, and nothing is lost by working with the simple identities of (\ref{eq:ProductIdentity2}) and the $\Phi$ fields only.

With the ground so prepared we can finally apply the semantic map $\hookrightarrow$ to the left hand side of the product identities to get

\begin{equation}\label{eq:ProductIdentityMapped1}
\sum_{{k=0,l=0}\atop \rm{cycl. perm.}}^{k+l=n}{\bf pr}(\Phi^k,{\bf pr}(\Phi^l)) \hookrightarrow
\sum_{{k=0,l=0}\atop \rm{cycl. perm.}}^{k+l=n}\langle\Phi|^{\otimes k}(\langle\Phi|^{\otimes l}|{\cal V}_{l+1}\rangle)\cdot|{\cal V}_{k+2}\rangle.
\end{equation}

In writing this equation, we are freely switching bra $\leftrightarrow$ ket Fock spaces as need arise to do the contractions. With $l=n-k$ we now have

\begin{equation}\label{eq:ProductIdentityMapped2}
\sum_{{k=0}\atop \rm{cycl. perm.}}^{n-1}\langle\Phi|^{\otimes k}\langle\Phi|^{\otimes n-k}|{\cal V}_{k+2}\rangle\cdot|{\cal V}_{n-k+1}\rangle=0.
\end{equation}

The sum stops at $k=n-1$ since the last term $k=n$ is zero, in the abstract product identities corresponding to ${\bf pr}(\,)=0$, which in the implementation would be $|{\cal V}_1\rangle=0$, i.e. there is no 1-vertex.  

Then focusing on the first ($k=0$) and next to last ($k=n-1$) terms, we see, using the conventions introduced for the 2-vertex, that they simply give us

\begin{equation}\label{eq:QVTerms}
\sum_{r=1}^{n+1}Q_r|{\cal V}_{n+1}\rangle.
\end{equation}

Then the rest of the terms in (\ref{eq:ProductIdentityMapped2}) pair off nicely in a similar way. Thus the ${\bf pr}(\Phi^k,({\bf pr}\Phi^{n-k}))$ term for $k\geq 1$ maps to precisely ${n\choose k}$ terms containing the vertex combination $|{\cal V}_{k+2}\rangle\cdot|{\cal V}_{n-k+1}\rangle$, while the ${\bf pr}(\Phi^{n-k-1},({\bf pr}\Phi^{k+1}))$ term maps to precisely ${n\choose n-k-1}$ terms also containing the vertex combination $|{\cal V}_{k+2}\rangle\cdot|{\cal V}_{n-k+1}\rangle$. Since ${n\choose k}+{n\choose n-k-1}={n+1\choose n-k}$ we see that we get precisely the correct number of terms to fully symmetrize $|{\cal V}_{k+2}\rangle\cdot|{\cal V}_{n-k+1}\rangle$. This is so because this contraction of vertices has $n+1$ free non-contracted indices (two of the $n+3$ being contracted). Doing the algebra carefully yields

\begin{equation}\label{eq:VVTerms1}
\sum_{k=1}^{\lfloor(n-1)/2\rfloor}|{\cal V}_{k+2}\rangle\diamond|{\cal V}_{n-k+1}\rangle,
\end{equation}

or, upon re-indexing the sum with $p=k-1$

\begin{equation}\label{eq:VVTerms2}
\sum_{p=0}^{\lfloor(n-3)/2\rfloor}|{\cal V}_{p+3}\rangle\diamond|{\cal V}_{n-p}\rangle.
\end{equation}

Hence, collecting all the terms, we get

\begin{equation}\label{eq:QVVVTerms}
\sum_{r=1}^{n+1}Q_r|{\cal V}_{n+1}\rangle=-\sum_{p=0}^{\lfloor(n-3)/2\rfloor}|{\cal V}_{p+3}\rangle\diamond|{\cal V}_{n-p}\rangle,
\end{equation}

which is exactly what we got before from the explicit $(S,S)=0$ calculation to all orders in $g$ and antighost number. 

In conclusion, the syntactically derived product identities of the sh-Lie algebra maps semantically to equations for the vertices in the Fock complex implementation. This result lends considerable strength to our framework.

%*****************************************
%DEFORMATION THEORY AND GAUGE INTERACTIONS
%***********************************************************
\subsection{Formal non-obstruction proof}\label{subsec:FNOP}
In \cite{BarnichHenneaux1993a} it was argued, based on homological perturbation theory, that there is no general "no-go" theorem for constructing consistent couplings for general gauge theories, at least not if locality is not an issue. If this is true it spells good news for higher spin gauge theory. However, since the relation between the antighost number and interaction order expansions were not entirely clear in that paper, we will briefly restate the argument in a way that is directly relevant to higher spin gauge field theory in the vertex deformation implementation. We will follow \cite{GomisParis1993} where the existence proof is framed in a power series expansion. It also provides a check on the efficiency of the formalism. 

We will not address the question of locality, even though spacetime locality is of course basic assumption in field theory. Interaction occurs locally in spacetime and disturbances propagate within the local lightcones. With respect to the number of derivatives in the higher spin vertices, locality is indeed an issue. The best we can say at the moment is that for a given interaction order $n$ and highest spin $s$, the number of derivatives is finite and of the order of ${\cal O}(ns)$. 

We want to show that the recursive equations (\ref{eq:S0SN}) issuing from the master equation can be solved and hence that $S_n$ exist to any order. First introduce the BRST differential ${\bf s}_0$ associated with the free part of $S$

\begin{equation}\label{eq:FreePartBSRT}
{\bf s}_0 X=(X,S_0).
\end{equation}

${\bf s}_0$ is nilpotent due to the nilpotency of $S_0$ and the Jacobi identity for the antibracket.

Next define the quantities $Q_n$ according to

\begin{equation}\label{eq:DefQN}
Q_n(S_{n-1},\ldots,S_1)={1\over 2}\sum_{k=1}^{n-1}(S_{k},S_{n-k})=0\quad\mbox{ for }n\geq2.
\end{equation}

Clearly then, the recursive relations (\ref{eq:S0SN}) read

\begin{equation}\label{eq:NewS0SN}
{\bf s}_0S_n+Q_n(S_{n-1},\ldots,S_1)=0\quad\mbox{ for }n\geq2.
\end{equation}

For the lowest order equation

\begin{equation}\label{eq:DeltaS1zero}
{\bf s}_0S_1=0\Leftrightarrow(S_0,S_1)=0,
\end{equation}

we will assume that there is a solution. Indeed, in the Fock complex vertex implementation we already have a solution in the form of the partial cubic vertex $|{\cal V}_3\rangle$ found in \cite{AKHB1988}. Although this vertex is not completely determined, the parts that are known can easily be seen as a truncation of the full vertex. We are confident that this vertex is correct as it reproduces Yang-Mills cubic interactions and furthermore is consistent with the light-front cubic vertices for all integer spin.

Thus the real work starts with showing that the equations (\ref{eq:NewS0SN}) starting with $n=2$ can be solved. As usual, there are two parts to the proof. First, $Q_n$ should be ${\bf s}_0$ closed, i.e.

\begin{equation}\label{eq:QNClosed}
{\bf s}_0Q_n=0.
\end{equation}

This is necessary since 

\begin{equation}\label{eq:QNClosedNec}
{\bf s}_0({\bf s}_0S_n+Q_n(S_{n-1},\ldots,S_1))=0\Rightarrow{\bf s}_0Q_n=0.
\end{equation}

Second, $Q_n$ should be ${\bf s}_0$ exact, i.e.

\begin{equation}\label{eq:QNExact}
Q_n={\bf s}_0X.
\end{equation}

This is sufficient, since then we can choose $X$ for $S_n$ modulo closed terms, or what in this context amounts to field redefinitions. 

Starting with the second part, $Q_n$ is ${\bf s}_0$ exact if the first cohomology group $H_1({\bf s}_0)$ is zero. As usual, we can prove that all cohomological groups $H_k({\bf s}_0)$ with $k\not=0$ are trivial, i.e. that ${\bf s}_0$ is acyclic in degree $k\not=0$, if we can exhibit a contracting homotopy $\sigma$ with the property 

\begin{equation}\label{eq:ContractinHomotopy}
\sigma{\bf s}_0+{\bf s}_0\sigma=N
\end{equation}

for some diagonalizable linear operator $N$. Rather than performing this construction here, we will lean against the literature were the triviality of the cohomology is proved, see for example \cite{HenneauxTeitelboim,GomisParis1993}. Instead, note that the recursive equations (\ref{eq:V3Eqs}) and (\ref{eq:SumQonNewVertex}) for the explicit vertices $|{\cal V}_n\rangle$, themselves trivialize the cohomology in that the master equation $(S,S)$ at any antighost number is implied by these equations. A proof that {\it these} equations can be solved will not be attempted here.

Turning then to the first part of the proof, this is a simple induction proof. For the base step, choose $n=2$. We get zero using (\ref{eq:DeltaS1zero})

\begin{equation}\label{eq:BaseStep}
{\bf s}_0Q_2=(S_1,{\bf s}_0S_1)=0.
\end{equation}

For the induction step, we assume the recursive equations (\ref{eq:NewS0SN}) to hold for all $i\leq n-1$. For $n$ we get using (\ref{eq:DefQN})

\begin{eqnarray}\label{eq:IndStep}
{\bf s}_0Q_n={1\over 2}\sum_{k=1}^{n-1}\big(({\bf s}_0S_{k},S_{n-k})-(S_{k},{\bf s}_0S_{n-k})\big)= \nonumber \\
{1\over 2}\sum_{k=1}^{n-1}\big((Q_k,S_{n-k})-(S_{k},Q_{n-k})\big)= \nonumber \\
\sum_{k=1}^{n-1}\big(Q_{k},S_{n-k})\big)= \nonumber \\
{1\over 2}\sum_{k=1}^{n-1}\sum_{l=1}^{k-1}((S_{l},S_{k-l}),S_{n-k})=\nonumber \\
{1\over 6}\sum_{k=1}^{n-1}\sum_{l=1}^{n-k-l}\Big[((S_{l},S_{k}),S_{n-k-l}))+\mbox{ cyclic permutations}\Big]=0.
\end{eqnarray}

We get zero by re-indexing and employing the Jacobi identity for the antibracket.

\pagebreak
%===================================================================================================================
%THE VERTEX IMPLEMENTATION
%===================================================================================================================
\section{The vertex implementation to all orders}\label{sec:VertexImplementation}
The vertex implementation provides an explicit ansatz for the interacting theory in terms of a formal power series. We have already used it extensively in section \ref{sec:DeformationTheory}} in an abstract way. Here we will elaborate it as far as possible without actually computing the coefficients. The starting point is the master action $S$ in equation (\ref{eq:VertexActionS1})

\begin{equation}\label{eq:VertexActionS2}
S=\langle\Psi|Q|\Psi\rangle|_{{\rm gh_m}=0}+\sum_{n=3}^\infty g^{n-2}\langle\Psi|^{\otimes n}|{\cal V}_n\rangle|_{{\rm gh_m}=0}.
\end{equation}

The vertex operators ${\cal V}_n$, to the extent that they can be determined explicitly, thus encode higher spin gauge interactions to all orders in formal perturbation theory. They also encode, as already discussed, gauge transformations and all higher order consistency requirements. The expansion is formal in the sense that we do not discuss convergence or summation issues.

We will proceed to state everything that is known about the general form of the vertex operators ${\cal V}_n$.

%*************************
%THE VERTEX IMPLEMENTATION
%**************************************
\subsection{The vertices to all orders}
We work in a momentum space representation of the fields. Thus each field $\Psi_{k}$ with $k\in\{1,\ldots,n\}$, entering the $n$-vertex, comes with its own momentum ${\bf p}_k$ and lives in its own numbered copy of the higher spin fiber of section \ref{subsec:ASHSF}. We use the notational conventions

\begin{equation}\label{eq:}
|-\rangle_{1\cdots n}=\prod_{i=1}^n\int d^4p_i|-\rangle_i \delta^4\Big(\Sigma_{i=1}^n p_i\Big),
\end{equation}

\begin{equation}\label{eq:}
|{\cal V}_n\rangle={\cal V}_n|-\rangle_{1\cdots n},
\end{equation}

\begin{equation}\label{eq:}
\langle\Psi|^{\otimes n}=\langle\Psi_1|\langle\Psi_2|\cdots\langle\Psi_n|.
\end{equation}

In order to enforce the restriction of $S$ to ${\rm gh_m}=0$, we restrict $\langle\Psi|^{\otimes n}$ to ${\rm gh_m}=-n/2$ (see section \ref{subsec:PCITVI}). By convention we chose the coupling constant to be dimensionless, or ${\rm d}(g)=0$. Then we get ${\rm d}({\cal V}_n)=D+n-nD/2$ which will be carried by a second coupling constant $\kappa$ with ${\rm d}(\kappa)=-1$. By dimensional arguments, no further coupling constants are needed, and it would be natural to identify $g$ with a Yang-Mills coupling constant, and $\kappa$ with a gravitational coupling constant. The spin $s$ coupling constant $g_s$ is then given by $g_s=g\kappa^{s-1}$. At the present stage of investigation these are purely formal identifications. Nevertheless, the framework set up here does not in itself demand the introduction of separate coupling constants $g_s$ for each spin $s$. 

%**********************************************
\subsubsection*{Ansatz for the vertex operator}
The ansatz for the vertex operator ${\cal V}_n$ is based on the following clauses
\begin{itemize}
\item ${\rm gh_m}({\cal V}_n)=0$,
\item ${\cal V}_n$ does not contain annihilators $c^0$, $c^-$, $b_+$ or $\alpha_\mu$,
\item ${\cal V}_n$ is a spacetime scalar,
\item ${\rm d}({\cal V}_n)={1\over 2}(2D+2n-nD)$.
\end{itemize}

The first three clauses imply that ${\cal V}_n$ can be built from the bilinear combinations $\alpha^\dagger_r\cdot\alpha^\dagger_s,\,\alpha^\dagger_r\cdot p_s,\, c^+_rb_{-s},\, c^+_rb_{0s},\, p_r\cdot p_s$, where the indices $r,s$ label HS fields. It is the fourth clause that requires us to introduce at least one dimensional constant $\kappa$ to balance the dimensions for the second, fourth and fifth bilinear (the fifth bilinear was inadvertently left out in \cite{AKHB2005a}). 

Introduce a symbol $\eta_{rs}^a$ to denote the dimensionless bilinears according to

\begin{equation}\label{eq:VertexBilinears}
\cases{\eta_{rs}^1=\alpha^\dagger_r\cdot\alpha^\dagger_s\cr\eta_{rs}^2=\kappa\alpha^\dagger_r\cdot p_s\cr\eta_{rs}^3=c^+_rb_{-s}\cr\eta_{rs}^4=\kappa c^+_rb_{0s}\cr\eta_{rs}^5=\kappa^2 p_r\cdot p_s.}
\end{equation}
\vskip 0.2cm

The higher spin vertices cannot be built out of these bilinear combinations alone if we want to reproduce Yang-Mills cubic couplings for spin 1 \cite{AKHB1988,KohOuvry}. Furthermore, from the light-front form of the cubic interaction term \cite{BBL1987} it is known the vertex must at least contain terms of the generic form $\alpha^{\dagger}\alpha^{\dagger}\alpha^{\dagger}p$ (indices suppressed), i.e. with three oscillators and one momentum label. Such a covariant vertex was partially determined in \cite{AKHB1988} and it correctly reproduces the Yang-Mills cubic interaction term. Thus powers of the $\eta$'s must be considered, and the generic form of the vertex is an exponential of a sum of such powers of $\eta$'s with appropriate numerical coefficients.

To be definite, introduce a symbol $\Delta_{2m}^n$ where $n$ denotes the order of the vertex and $m$ denotes the homogenous power of bilinears,

\begin{equation}\label{eq:VertexDeltas}
\Delta_{2m}^n={\sp n}Y^{r_1s_1\cdots r_ms_m}_{a_1\cdots a_m}\eta_{r_1s_1}^{a_1}\cdots\eta_{r_ms_m}^{a_m},
\end{equation}
where there are implicit summations according to
\begin{itemize}
\item all $r_i$ and $s_i$, $i\in[1..m]$ are summed over the list $[1..n]$,
\item all $a_i$, $i\in[1..m]$ are summed over the list $[1..5]$,
\end{itemize}
and where the coefficients ${\sp n}Y^{r_1s_1\cdots r_ms_m}_{a_1\cdots a_m}$ are algebraic numbers to be determined.

Finally, ${\cal V}_n$ can be synthesized as
\begin{equation}\label{eq:}
{\cal V}_n=\kappa^{({nD\over 2}-D-n)}\exp\Big(\sum_m^\infty\Delta_{2m}^n\Big).
\end{equation}

In this framework, the countable set of operators $\{{\cal V}_n\}_{n=3}^\infty$, if they exist, encode the full interacting theory of higher spin gauge fields. The same information can therefore also be considered as encoded into the countable set of numbers $\{{\sp n}Y^{r_1s_1\cdots r_ms_m}_{a_1\cdots a_m}\}$. 

There is one puzzle here that will eventually have to be solved by explicit calculation. From the partial determination of the covariant cubic vertex in \cite{AKHB1988} we know, as already noted, that cubic Yang-Mills can be reproduced by terms in the vertex of the form ${\sp 3}Y^{rstu}_{12}\eta_{rs}^1\eta_{tu}^2={\sp 3}Y^{rstu}_{12}\alpha^\dagger_r\cdot\alpha^\dagger_s\alpha^\dagger_t\cdot p_u$ or simply $Y^{rstu}\alpha^\dagger_r\cdot\alpha^\dagger_s\alpha^\dagger_t\cdot p_u$ in the notation of \cite{AKHB1988}. The same generic form $\alpha^\dagger\alpha^\dagger\alpha^\dagger p$ also follows from the light-front analysis of \cite{BBL1987}. Furthermore, on the light-front all higher spin cubic vertices results from this very term by just extracting the successive powers of $\alpha^\dagger\alpha^\dagger\alpha^\dagger p$ in $\exp (Y\alpha^\dagger\alpha^\dagger\alpha^\dagger p)$. Whether this ''factorization'' property holds also covariantly is not known, but can of course be settled by direct calculation. Otherwise, we will need leading terms of the generic form ${\sp 3}Y^{r_1s_1r_2s_2r_3s_3r_4s_4}_{1122}\alpha^\dagger_{r_1}\cdot\alpha^\dagger_{s_1}\alpha^\dagger_{r_2}\cdot\alpha^\dagger_{s_2}\alpha^\dagger_{r_3}\cdot p_{s_3}\alpha^\dagger_{r_4}\cdot p_{s_4}$ and similarly for higher spin. This is also linked to the question of field redefinitions. 

%*************************
%THE VERTEX IMPLEMENTATION
%************************************************
\subsection{Field redefinitions}\label{subsec:FR}
Field redefinitions are non-linear transformations of the fields which when inserted into the free action produces fake interactions in the form of non-linear terms \cite{BerendsBurgersvanDam1985}. Hence, the true interactions must be constructed modulo non-linear terms that can be removed by non-linear field transformations. For a generic field we would have

\begin{equation}
\varphi\rightarrow\varphi_r=\varphi+gR_3(\varphi,\varphi)+g^2R_4(\varphi,\varphi,\varphi)+\ldots.
\end{equation}

The general form of the non-linear terms are governed by dimensionality and index structure. Thus for a generic spin $s$ field $\varphi^{(s)}$, the form of the quadratic term is

\begin{equation}
\varphi^{(s)}_r=\varphi^{(s)}+g_sR_3(\partial^{(s-2)}\varphi^{(s)}\varphi^{(s)}),
\end{equation}

where the number of derivatives are determined by the dimensionality count which works out as $(1-s)+(s-2)+1+1=1$. Then the number $x$ of index contractions can be calculated by balancing the number of free indices on both sides of the equation $s=(s-2)+s+s-2x\Rightarrow x=s-1$.

Generalizing these arguments, we introduce field redefinition vertices $|{\cal R}_n\rangle$ and write 

\begin{equation}
|\Psi\rangle\rightarrow|\Psi_r\rangle =|\Psi\rangle +\sum_{n=2}^\infty g^{(n-1)}\langle\Psi|^{\otimes n}|{\cal R}_{n+1}\rangle,
\end{equation}

where, as usual, antighost number matching is implicit. Performing field transformations of this form on the free action produce fake interaction terms of the form

\begin{equation}
\sum_{n=3}^\infty g^{(n-2)}\langle\Psi|^{\otimes n}\sum_{r=1}^nQ_r|{\cal R}_n\rangle.
\end{equation}

Comparing to the general form of the interactions given in equation (\ref{eq:VertexActionS2}), we see that fake interactions can be characterized by

\begin{equation}
|{\cal V}_n\rangle_{\rm fake}=\sum_{r=1}^nQ_r|{\cal R}_n\rangle,
\end{equation}

which in the language of homology is to say that fake interaction vertices are exact.

%*************************
%THE VERTEX IMPLEMENTATION
%***************************************************************************
\subsection{Global symmetries and subsidiary conditions}\label{subsec:GSASC}
The free field theory actually supports some more structure that we will now discuss. Consider the set of operators

\begin{equation}\label{eq:SetOfOpeartors}
\cases{T={1\over 2}\alpha\cdot\alpha+b_+c^-\cr 
T^\dagger={1\over 2}\alpha^\dagger\cdot\alpha^\dagger-b_-c^+\cr 
N={1\over 2}(\alpha\cdot\alpha^\dagger+\alpha^\dagger\cdot\alpha)-b_+c^+ +b_-c^-\cr 
P=\alpha\cdot p-2b_0c^-\cr 
P^\dagger=\alpha^\dagger\cdot p+2b_0c^+\cr
G=p^2.}
\end{equation}
\vskip 0.2cm

All these operators commute with the BRST charge $Q$. The first three together span an $SU(1,1)$ algebra with

\begin{equation}\label{eq:}
[T,T^\dagger]=N,\quad [N,T]=-2T,\quad [N,T^\dagger]=2T^\dagger.
\end{equation}

The last three together span a copy of the underlying mechanics gauge algebra

\begin{equation}\label{eq:}
[P,P^\dagger]=G,\quad [G,P]=0,\quad [G,P^\dagger]=0.
\end{equation}

From these operators we can form hermitean combinations ${\cal P}=P+P^\dagger$ and ${\cal T}=T+T^\dagger+N$ that commute among themselves and with $Q$. 

It is natural to consider these operators as generating global symmetries of the free theory according to 

\begin{eqnarray}
\delta_{\cal P}|\Psi\rangle=i\epsilon{\cal P}|\Psi\rangle,\label{eq:GlobalP}\\ 
\delta_{\cal T}|\Psi\rangle=i\epsilon{\cal T}|\Psi\rangle.\label{eq:GlobalT}
\end{eqnarray}

It then follows  that $\delta_{\cal P}\langle\Psi|Q|\Psi\rangle=\delta_{\cal T}\langle\Psi|Q|\Psi\rangle=0$. Extending these global symmetries to the interacting theory yield the following conditions on the vertices

\begin{eqnarray}
\sum_{r=0}^n{\cal P}_r|{\cal V}_n\rangle=0,\label{eq:GlobalPonV}\\ 
\sum_{r=0}^n{\cal T}_r|{\cal V}_n\rangle=0.\label{eq:GlobalTonV}
\end{eqnarray}

Both of these equations were used in \cite{AKHB1988} when determining parts of the cubic vertex. The first equation (\ref{eq:GlobalPonV}) was identified as a global symmetry, but the origin of the second (\ref{eq:GlobalTonV}) was unclear in that paper.

The presence of these global symmetries is an effect of the auxiliary fields contained in the components $F$ and $H$ in $|\Phi\rangle$. Modulo the field equations, the ${\cal P}$ symmetry reduces to rigid gauge transformations, as can be surmised from the form of the operators $P$ and $P^\dagger$. The ${\cal T}$ symmetry is somewhat more obscure. It is related to the awkward tracelessness conditions on the fields and gauge parameters. In the next section we will argue that we can actually discard these constraints, and that it is natural to do so when working with the higher spin fiber $|\Phi\rangle$ subject to the global transformations (\ref{eq:GlobalTonV}).

%*************************
%THE VERTEX IMPLEMENTATION
%*********************************************************************
\subsection{Discarding the tracelessness conditions}\label{subsec:DTC}
As is well known, the tracelessness conditions on the higher spin gauge fields and gauge parameters are needed in order to ensure that the correct number of degrees of freedom are propagated. Let us remind ourselves how it works in four dimensions \cite{deWitFreedman}. In $D=4$, a symmetric tensor with $s$ indices has ${s+3\choose 3}$ components. We also know that the number of physical helicity states for a massless field of arbitrary spin $s\geq 1$ is precisely 2 in $D=4$. Thus for spin 1, the four components of $A_\mu$, carry two states of helicity when the  gauge (and re-gauge) transformations $\partial_\mu\xi$ are taken into account, i.e. $4-2\cdot 1=2$. Likewise for spin 2, the ten components of $A_{\mu\nu}$, carry two states of helicity when the transformations $\partial_{(\mu}\xi_{\nu)}$ are taken into account, i.e. $10-2\cdot 4=2$. Starting with spin 3, the gauge parameters must be traceless $\xi^{\,\prime}_\mu=0$, and starting with spin 4, the fields must be double traceless $A^{\prime\prime}=0$. Thus the spin 3 and 4 counting look like $20-2\cdot(10-1)=2$ and $35-2\cdot(20-4)-1=2$ respectively. This generalizes to 

\begin{equation}\label{eq:DFCount1}
{s+3\choose 3}-2\Bigg({s+2\choose 3}-{s\choose 3}\Bigg)-{s-1\choose 3}=2
\end{equation}

for all $s\geq 4$.

In the higher spin fiber, these trace constraints are implemented using the $T$ operators simply by demanding $T|\Phi\rangle=T|\Xi\rangle=0$ in the BRST-gauge formulation and by $T|\Psi\rangle=0$ in the BV formulation. Either way, constraints like these are awkward to work with, and it would be interesting to find a way to do without them. The physical reason why they are present in the first place can be discerned by simple degrees of freedom counting. Starting with spin 3 and requiring neither $\xi^{\,\prime}=0$ nor $A^{\prime\prime}=0$, we get the count

\begin{equation}\label{eq:DFCount2}
{s+3\choose 3}-2{s+2\choose 3}={1\over 6}(6+7s-s^3),
\end{equation}

which is 0 for spin 3 and then goes negative for higher spin. Hence the gauge parameters must be constrained in order not to remove to many degrees of freedom.

Starting with spin 4 and requiring $\xi^{\,\prime}=0$ but not $A^{\prime\prime}=0$, we instead get the count

\begin{equation}\label{eq:DFCount3}
{s+3\choose 3}-2\Bigg({s+2\choose 3}-{s\choose 3}\Bigg)={1\over 6}(6 + 11s - 6s^2 + s^3),
\end{equation}

which correctly yields 2 for spin 3, but then starts to give too many degrees of freedom from spin 4 onwards. These are of course precisely killed off by the number of degrees of freedom 

\begin{equation}\label{eq:DFCount4}
{s-1\choose 3}={1\over 6}(-6+11s-6s^2+s^3),
\end{equation}

removed by double tracelessness condition. Subtracting this number of states restores the balance to precisely 2 degrees of freedom. The double tracelessness constraint can be understood as preventing lower spin components of the symmetric tensor fields from propagating. This makes sense in a pure spin $s$ theory, but in a theory already containing fields of all integer spin, there is no reason to exclude lower spin components as long as they do not ruin consistency.

Let us now turn to the higher spin fiber $|\Phi\rangle$. It contains apart from the higher spin fields $A^{(s)}$ ($s$ symmetrized indices) also concomitant auxiliary fields $H^{(s-1)}$ and $F^{(s-2)}$. Of these, the $H$ fields are unproblematic as their equations of motion are algebraic (for notation, see appendix) 

\begin{equation}
H^{(s-1)}={1\over 2}(-)^s\Big(s\partial\cdot A^{(s-1)}-{1\over (s-1)}\partial^{(1}F^{s-2)}\Big),
\end{equation}

and they can be easily solved for if required. Their presence in the theory is related to the global ${\cal P}$ symmetry. They can be neglected when counting degrees of freedom as they are clearly not independent. 

The $F$ fields however furnish us with independent degrees of freedom as can be seen from their equations of motion

\begin{equation}
\partial^2 F^{(s-2)}+2(-)^{(s+1)}(s-1)\partial\cdot H^{(s-2)}=0.
\end{equation}

As already noted, requiring $T|\Xi\rangle=0$ yields $\xi^{\,\prime}=0$ for all $s\geq 3$, while requiring $T|\Phi\rangle=0$ yields

\begin{eqnarray}
F^{(s-2)}={s\choose 2}A^{\prime(s-2)}, \\
F^{\,\prime (s-4)}=0.
\end{eqnarray}

These equations then allow us to solve for $F$ while at the same time obtain $A^{\prime\prime}=0$ for $s\geq 4$, and we would be back in the standard formulation using just $A$ fields.

The question is now, does the theory make sense if we keep the $F$ fields and at the same time discard the trace constraints on fields and parameters? In order to sketch an answer to that question, first of all note that while the $F$ fields are independent degrees of freedom, they suffer no independent gauge transformations. Indeed we have

\begin{eqnarray}\label{eq:}
\delta A^{(s)}={1\over s}\partial^{(1}\xi^{s-1)},\\
\delta F^{(s-2)}=(s-1)\partial\cdot\xi^{(s-2)}.
\end{eqnarray}

Let us then repeat the degrees of freedom count. Adding the number of states in the $A$ and $F$ fields and subtracting the gauge and re-gauge states, we get

\begin{equation}\label{eq:DFCount5}
{s+3\choose 3}+{s+1\choose 3}-2\cdot{s+2\choose 3}=s+1.
\end{equation}

This is of course correct for spin 0 and 1. For spin 2 we would get an extra propagating scalar field, precisely the scalar field $F$ that would otherwise be equated with the trace $A^{\prime}$ of the metric perturbation field $A_{\mu\nu}$. For spin 3, the interpretation would be an extra propagating spin 1 field $F_\mu$. 

The general case for odd spin $s$ would be propagating fields of spin  $s,s-2,\ldots ,1$, each one contributing 2 physical degrees of freedom, making up the sum $2+2+\ldots +2=s+1$. Likewise, for even spin $s$, there would be propagating fields of spin $s,s-2,\ldots 2,0$, each one contributing 2 physical degrees of freedom, except the last scalar, making up the sum $2+2+\ldots +2+1=s+1$. Thus it could be expected that number of physical propagating lower spin components of the fields $A^{(s)}$ and $F^{(s-2)}$ are precisely correct. Although the fact that the degrees of freedom count works out nicely lends support to the consistency of the theory, a detailed analysis of the field equations is needed to confirm it. Although we have only shown the counting in the case of $D=4$, it goes through, for free fields, in higher dimensions as well.

Returning now to the case discussed above with traceless gauge parameters, but with no double tracelessness constraint on the $A$-fields and no $F$-fields, corresponding to equation (\ref{eq:DFCount3}), we can see the difference. In that case there are not enough gauge transformations to gauge away unphysical (negative metric) states.  

One immediate consequence of this is the appearance of extra ''low'' spin fields, indeed, there will be an infinite number of spin 1 fields, spin 2 fields et cetera, as every higher spin field will come with a full spectrum of lower spin components. However, as ''pure'' spin 3 fields for instance interacts cubically with three derivatives, the concomitant spin 1 fields will also have higher derivative interactions, thus not being in obvious conflict with the finite number of observed spin 1 fields in nature. The same argument holds for spin 2 and there will be just one graviton. Presumably this also explains recent results on higher than minimal number of derivatives in cubic spin 3 interactions \cite{BekaertBoulangerCnockaert2006a}, in particular with five derivatives which corresponds precisely to the spin 3 component of the spin 5 level in the fiber. It is natural to conjecture that all such interactions are contained in the vertices $|{\cal V}_n\rangle$ as there are no restrictions on these vertices other than consistency of the gauge structure and the global symmetries.

Discarding the tracelessness constraints on the gauge transformations also has the additional benefit of alleviating any doubt as to the irreducibility of the theory.

In conclusion then, we can now understand how the auxiliary fields of the higher spin fiber are related to the global symmetries. In the presence of the $H$ fields we have ${\cal P}$ symmetry. Likewise, in the presence of the $F$ fields we have ${\cal T}$ symmetry.

\pagebreak
%===================================================================================================================
%CATEGORIES
%===================================================================================================================
\section{Elements of a formulation in terms of categories}\label{sec:Categories}
Having so worked some distance into the nitty-gritty details of an explicit implementation, we will end by gazing in the direction of further abstraction and begin a treatment of higher spin gauge field theory using the tools of category theory. As such it is very close in spirit to the computer science inspired abstract interface formalism. Indeed, categories can be invoked to understand questions of syntax and semantics in computer science, and this is precisely the borderline we try to straddle. 

Using categories in field theory has been advocated in particular by J. Baez \cite{Baez2004a}, and since this is clearly a huge area of research, we will confine ourselves to the mere outlines of such a treatment in the particular case of higher spin theory. Thus we warn the reader that what follows is ''heuristic'' and serves mainly as a guide to future research.

%**********
%CATEGORIES
%********************************************
\subsection{The abstract interface revisited}
Before proceeding, let us return to the abstract interface and adapt it to the BRST-BV approach. Thus instead of the type $\Phi::{\cal H}$ we work with a (ghost number graded) type $\Psi::{\cal H}$. The abstract action becomes

\begin{equation}\label{eq:AbstractS1}
S=\sum_{i=2}^\infty{g^{i-2}\over i!}\sum_{\pi[i]}{\bf vx}(\Psi_1,\Psi_2,\ldots,\Psi_i)=\sum_{\pi[i=2]}{g^{i-2}\over i!}{\bf vx}(\Psi^i),
\end{equation}

or written in terms of the inner product ${\bf in}$ and the products ${\bf pr}$

\begin{equation}\label{eq:AbstractS2}
S=\sum_{\pi[i=2]}{g^{i-2}\over i!}{\bf in}(\Psi,{\bf pr}(\Psi^{i-1})).
\end{equation}

We take it for implicitly understood that the sums should be evaluated at ghost number ${\rm gh_f}=0$. Expanding according to antighost number ${\rm gh_a}$ yields for the first two levels

\begin{eqnarray}\label{eq:AbstractSLevels01}
S^{(0)}&=&\sum_{\pi[i=2]}{g^{i-2}\over i!}{\bf in}(\Phi,{\bf pr}(\Phi^{i-1}))\\
S^{(1)}&=&\sum_{\pi[i=2]}{g^{i-2}\over i!}{\bf in}(\Phi^\#,{\bf pr}({\cal C},\Phi^{i-2})).
\end{eqnarray}

In order two calculate the first term $(S^{(0)},S^{(1)})$ in $(S,S)$ we just need to know how the abstract antibracket $(\Psi,\Psi^\#)=1$ acts on the abstract inner product ${\bf in}(X,Y)$. There is really just one reasonable way to define this. Suppose that we know $(X,Y)=0$. In this special case it is natural to define

\begin{equation}\label{eq:InnerProdBracketRel}
({\bf in}(\Psi,X),{\bf in}(\Psi^\#,Y))={\bf in}(X,Y).
\end{equation}

This brief summary shows that in order to categorify higher spin gauge theory, there are three distinct operations that need to be formalized: the products ${\bf pr}$ (this is the easy part), the inner product ${\bf in}$ and the bracket $(\,,)$. We will just briefly discuss the first one.

%**********
%CATEGORIES
%*************************************************
\subsection{Monoidal categories of gauge fields}
There is a considerable freedom of choice when casting higher spin gauge field theory in the language of category theory. Partly this has to do with the versatility of category theory itself, but also with field theory physics. To fix ideas, consider a scalar field $\varphi(x)$. Implicitly we think of this field as permeating a region of spacetime, so there is in a sense {\it just one} field, and we would write a local self interaction as a power of the field, say $\varphi(x)^4$. However, transforming to momentum space we would write the interaction as

$$
\varphi(p_1)\varphi(p_2)\varphi(p_3)\varphi(p_4)\delta(p_1+p_2+p_3+p_4).
$$

Now it makes sense to think of this as an interaction between four objects, indeed a collision between four scalar particles. Which physical picture is most appropriate? Being pictures, both of them is the only reasonable answer. They simply focus different aspects of field theory.

As already discussed in section \ref{sec:Fields}, the mathematical notation $\varphi(x)$ lends itself to various connotations, and the same is even more pronounced for the abstract HS field $\Psi(\sigma)$ since we are then not restricted to a configuration space vs. momentum space set up. In order to discuss interactions in terms of $n$-ary field products ${\bf pr}$ we must however provide handles $\sigma_i$ to the fields and we are lead to a momentum space resembling notation. What we have is indeed $n$-ary field configurations $\{\Psi(\sigma_i)\}_{i=1}^n$. This is precisely the point where we have a choice of categorification \cite{BaezDolan1998a}.

One way would be to chose as objects of the category precisely $n$-ary field configurations $\{\Psi(\sigma_i)\}_{i=1}^n$. The arrows are then naturally defined as morphism taking any $n$-ary field configuration $\{\Psi(\sigma_i)\}_{i=1}^n$ to any $m$-ary field configuration $\{\Psi(\sigma_i)\}_{i=1}^m$. Since such an $n\rightarrow m$ morphism would connect $n+m$ individual fields $\Psi(\sigma_i)$, it would correspond to a $(n+m)$-point interaction. The physical picture is that of Feynman diagrams with $n$ in-fields and $m$ out-fields. In order for arrows like these to satisfy the category axioms, an $n\rightarrow m$ morphism can only be composed with an $p\rightarrow q$ morphism if $m=p$ corresponding to the need for the co-domains and domains of the composed morphisms to match. We must also allow disconnected morphisms. Of course, the morphism are more general than the basic classical interaction vertices which only form a subset of all interactions. Or put somewhat differently, composing the morphisms of the basic interaction vertices will produce Feynman diagrams which themselves are morphisms. The question of actually computing such diagrams leads straight into quantum theory and is clearly out-of-bounds at present.

However, the distinction between in-fields and out-fields is not very significant, so there is a certain amount of redundancy in the above set of $n\rightarrow m$ morphisms. A slightly different picture ensues in we consider instead a category consisting of labeled objects $\Psi(\sigma_i)$, and instead of arrows, we consider multi-arrows taking $n$ input objects, producing a new output object. These arrows can be thought of as $n$-ary operations. What we get then could be described as a {\it multicategory} \cite{Leinster}. Still a further simplification can be contemplated.

Although our objects $\Psi(\sigma_i)$ are labeled by $\sigma$, they do not really belong to different types if we choose to consider the ghost number grading to be internal to the type. Taking stock of this, we can consider a monoidal category with just one object $\Psi(\sigma)$. This might seem like a very strange notion at first. What can possibly be done with just one object? Consider however the following analogy. Take the category of vector spaces ${\bf Vect}$. This is a huge category, containing as it does not just Euclidean spaces of any dimension, but also all kinds of function spaces. Picking one of all these objects, say three-dimensional Euclidean space, we still have a rich structure to do mathematics in. This is the proper analogy. We are not really interested in the category of all things that could possibly be called ''fields''. We need a certain type of gauge fields, all depending on the implementation of the abstract structure. This kind of multicategory is precisely an {\it operad} \cite{May1997}, operads being mathematical structures formalizing $n$-input $\rightarrow1$-output operations. 

So in concluding this introduction to further development, it is quite clear that we need to devise some kind of monoidal category, ${\bf HSField}$ say, with a multi-operation corresponding to the multi-products. The monoidality just means that we put the extra structure of tensor products on the category, which is clearly needed in order to define the products ${\bf pr}::{\cal H}^{\otimes n}\rightarrow{\cal H}$. We have a choice as to formalize in terms of operads, or in terms of multicategories (if we consider the different ghost number components of ${\Psi}$ as different types) or in terms of {\it PROPs} (corresponding to the $n\rightarrow m$ morphisms above).

%**********
%CATEGORIES
\subsection{Conclusions and a conjecture concerning quantization}
We have seen that the computer science inspired approach of the present paper and the preceeding one, working on the one hand with the pure syntax of the theory, and on the other, with an explicit concrete implementation, has gone a long way towards controlling the inherent complexity of the problem. Rephrasing all this in terms of categories will not just sharpen the formalism itself. It will also provide a natural setting, in terms of a properly defined functor, for the semantic map taking the abstract sh-Lie structure to the recursive (cohomological) equations governing the concrete vertices. This isomorphism between the higher homotopy and BRST-BV cohomology aspects of the theory, can presumably be exploited not just for explicitly constructing the vertices, but also in an attempt to quantize the theory.

There is a quite clear drift in the arguments. Suppose that we had explicit expressions for the higher spin interaction terms in the action. Formally quantizing the theory would yield vertices in the sense of perturbative quantum field theory. It is then not a great leap of imagination to think of generalized propagators for the fields $\Phi$. Stringing vertices together with propagators, we simply get Feynman diagrams. These would look just as spin network graphs with arbitrarily high order nodes. We could then conjecture that quantum higher spin gauge field theory produces generalized spin networks with edges corresponding to infinite multiplets of representations of a (spacetime) group $G$. The nodes are intertwining operators between these representations. The intertwining operators fall into the structure of a strongly homotopy algebra.

\pagebreak
\section{Appendix}
\subsection{Properties of the antibracket}
For definiteness, we work in a graded vector space (really an algebra) of fields $\Psi\supset\{\ldots,C_\alpha,\varphi_i,\varphi^\#_i,C^\#_\alpha,\ldots\}$ (as in section \ref{sec:BRSTBV}) defined on a spacetime manifold $\bf M$. The antibracket, defined in (\ref{eq:AntiBracket}) is such that

\begin{equation}\label{eq:FieldAntiFieldBracket2}
(\Psi^A(x),\Psi^\#_B(y))=\delta^A_B\delta(x-y).
\end{equation}

The bracket has the following properties for generic fields $X,Y,Z$

\begin{eqnarray}
\rho(X,Y)&=&\rho_X+\rho_Y+1 \quad({\rm mod}\;2)\;\;\;\quad\mbox{Grassman parity}\\
{\rm gh_f}(X,Y)&=&{\rm gh_f}(X)+{\rm gh_f}(Y)+1 \;\;\,\quad\quad\mbox{Ghost number = 1}\\
(X,Y)&=&-(-1)^{(\rho_X+1)(\rho_Y+1)}(Y,X) \quad\mbox{Symmetry}\\
(X,YZ)&=&(X,Y)Z+(-1)^{\rho_Y(\rho_X+1)}Y(X,Z) \quad\quad\mbox{Derivation}\\
(X,(Y,Z))&=&((X,Y),Z)+(-1)^{(\rho_X+1)(\rho_Y+1)}(Y,(X,Z))\;\mbox{Jacobi}
\end{eqnarray}

Note that the bracket $(\,\cdot)$ itself is odd.

This structure, $(\cdot,\,\cdot)::\Psi\otimes\Psi\rightarrow\Psi$, is sometimes called an Gerstenhaber algebra in the mathematics literature.

\subsection{Condensed notation for gauge fields}
The following condensed notation for symmetric tensor fields is fairly standard. A symmetric tensor with $n$ indices $\varphi_{\mu_1\ldots\mu_n}$ is written $\varphi^{(n)}$. Where, by the way, we do not care to distinguish upper and lower indices. Explicit symmetrization is done with weight 1, for instance $\partial_{\mu_1}\xi_{\mu_2\cdots\mu_n}+\ldots+\partial_{\mu_n}\xi_{\mu_1\cdots\mu_{n-1}}$ is written as $\partial^{(1}\xi^{n-1)}$. A dot $\cdot$ signifies a trace, and a double dot $:$ signifies a double trace. The traced tensor is decorated with a ''prime'' or a '' double prime'' and the number of remaining symmetrized indices. Thus tracing an $n$ index tensor $\varphi^{(n)}$ is written as $\varphi^{\prime(n-2)}$. Likewise, dots and double dots are also used for divergencies and double divergencies as in $\partial\cdot\varphi^{(n-1)}$. If care is exercised, it is actually possible to calculate with some reliability using this notation.

For the record, let us complement section \ref{subsec:DTC} by writing down the field equation for the spin $s$ gauge field $A^{(s)}$

\begin{equation}
\partial^2A^{(s)}+{2\over s}(-)^{(s+1)}\partial^{\,(1}H^{s-1)}=0.
\end{equation}

This equation together with the other field equations, and the tracelessness constraints, yield the Fronsdal field equations

\begin{equation}
\partial^2A^{(s)}-\partial^{\,(1}\partial\cdot A^{s-2)}+\partial^{(2}A^{\prime\,s-2)}=0.
\end{equation}

\subsection{Some conventions}
All low level calculations are done using conventions adopted from Weinberg \cite{WeinbergQFT1}, for instance, with space-like metric $\eta=(-+\cdots+)$ and $p_\mu=-i\partial_\mu$. A scalar field Lagrangian should therefore start out with ${1\over 2}\varphi\partial^{\,2}\varphi$.

Note that in $p$-space, a bra field $\langle A|$ comes with an opposite sign for the momentum. Thus $\langle A|$ is really shorthand for $\langle A(-p)|$ just as $|A\rangle$ is shorthand for $| A(p)\rangle$. A Fock space inner product (or contraction between Fock spaces) also involve momentum integration according to

\begin{equation}
\langle A|B\rangle=\int dp\langle A(-p)|B(p)\rangle
\end{equation}

When expanding the out the gauge fields as in 

\begin{equation}
\varphi=\varphi_0+\varphi^\mu\alpha^\dagger_\mu+\varphi^{\mu\nu}\alpha^\dagger_\mu\alpha^\dagger_\nu+\ldots
\end{equation}

we are really working in $p$-space. In $x$-space, the terms for odd spin should be multiplied by an $i$.

\pagebreak
\bibliographystyle{C:/TexDocs/BibStyle/unsrt}
\end{document}